\documentclass[%
 aip,
 amsmath,amssymb,
reprint,%
]{revtex4-1}

\usepackage{graphicx}% Include figure files
\usepackage{dcolumn}% Align table columns on decimal point
\usepackage{mathptmx}
\usepackage[utf8]{inputenc} % allow utf-8 input
\usepackage[T1]{fontenc}    % use 8-bit T1 fonts
\usepackage{hyperref}       % hyperlinks
\usepackage{booktabs}       % professional-quality tables
\usepackage{amsfonts}       % blackboard math symbols
\usepackage{nicefrac}       % compact symbols for 1/2, etc.
\usepackage{microtype}      % microtypography
\usepackage{subfigure}
\usepackage{multirow}
\usepackage{makecell}
\begin{document}

\preprint{AIP/123-QED}

\title{Discovering explicit Reynolds-averaged turbulence closures for turbulent separated flows through deep
learning-based symbolic regression with non-linear corrections}

\author{Hongwei Tang}
\affiliation{Jiangsu Key Laboratory of Hi-Tech Research for Wind Turbine Design,
Nanjing University of Aeronautics and Astronautics, 29 Yudao St., Nanjing 210016, China}

\author{Yan Wang}
\email{aerowangy@nuaa.edu.cn}
\affiliation{Jiangsu Key Laboratory of Hi-Tech Research for Wind Turbine Design,
Nanjing University of Aeronautics and Astronautics, 29 Yudao St., Nanjing 210016, China}
\affiliation{State Key Laboratory of Mechanics and Control of Mechanical Structures,
Nanjing University of Aeronautics and Astronautics, 29 Yudao St., Nanjing 210016, China}

\author{Tongguang Wang}
\affiliation{Jiangsu Key Laboratory of Hi-Tech Research for Wind Turbine Design,
Nanjing University of Aeronautics and Astronautics, 29 Yudao St., Nanjing 210016, China}

\author{Linlin Tian}
\affiliation{Jiangsu Key Laboratory of Hi-Tech Research for Wind Turbine Design,
Nanjing University of Aeronautics and Astronautics, 29 Yudao St., Nanjing 210016, China}


\begin{abstract}
This work introduces a novel data-driven framework to formulate explicit algebraic Reynolds-averaged Navier-Stokes
(RANS) turbulence closures. Recent years have witnessed a blossom in applying machine learning (ML) methods to
revolutionize the paradigm of turbulence modeling. However, due to the black-box essence of most ML methods, it is
currently hard to extract interpretable information and knowledge from data-driven models. To address this critical
limitation, this work leverages deep learning with symbolic regression methods to discover hidden governing equations of
Reynolds stress models. Specifically, the Reynolds stress tensor is decomposed into linear and non-linear parts. While
the linear part is taken as the regular linear eddy viscosity model, a long short-term memory neural network is employed
to generate symbolic terms on which tractable mathematical expressions for the non-linear counterpart are built. A novel
reinforcement learning algorithm is employed to train the neural network to produce best-fitted symbolic expressions.
Within the proposed framework, the Reynolds stress closure is explicitly expressed in algebraic forms, thus allowing for
direct functional inference. On the other hand, the Galilean and rotational invariance are craftily respected by
constructing the training feature space with independent invariants and tensor basis functions. The performance of the
present methodology is validated through numerical simulations of three different canonical flows that deviate in
geometrical configurations. The results demonstrate promising accuracy improvements over traditional RANS models,
showing the generalization ability of the proposed method. Moreover, with the given explicit model equations, it can be
easier to interpret the influence of input features on generated models.
\end{abstract}

\maketitle

\section{Introduction}\label{sec:intro}

Accurate predictions of turbulent flows are of paramount importance for many engineering applications of computational
fluid dynamics (CFD). With the increasing availability of computational resources over the last two decades,
scale-resolving simulations such as large eddy simulation (LES) and direct numerical simulation (DNS) have gained
widespread applications. While these methods indeed provide plenty of detailed insights into fluid flow physics, an
immediate difficulty relates to the computational cost, which could be prohibitively large for many practical
applications in physics and engineering sciences. By contrast, the computationally lower cost and superior robustness
leave the Reynolds-averaged Navier-Stokes (RANS) models still the most widely used tool in industrial simulations. This
scenario will remain unchanged into the coming decades \citep{doi:10.2514/1.J059953}. However, the predictive accuracy
of RANS equations could be severely plagued with the unsolved closure problem deeply rooted in RANS governing equations.
Such deficiency is strikingly serious for flow with complex geometries and large separations
\citep{wilcox1998turbulence}. Since the pioneering work of Boussinesq to develop a mathematical description of Reynolds
stress by introducing the concept of eddy viscosity, there has been a continuous attempt to formulate more accurate RANS
turbulence models.

In recent years, thanks to the rapid advancement of high-performance computing architectures and the increasingly
accessible high-fidelity flow data, the emerging machine learning (ML) and data science techniques provide a new
alternative in the analysis and understanding of turbulent flows \citep{doi:10.1146/annurev-fluid-010518-040547,
PhysRevFluids.6.050504}. This paradigm has been blossoming in turbulence modeling research. For example,
\citet{CHEUNG20111137} applied Bayesian techniques to calibrate the parameters of a traditional turbulence model against
experimental data. \citet{doi:10.1063/5.0097438} took the ML method to augment a turbulence model to enhance its
prediction in separation flow. These works try to improve the turbulence model performance without breaking the
Boussinesq assumption.

Rather than correcting the governing equations of existing turbulence models, another research subject is constructing
new constitutive models for Reynolds stress tensors. The related ideas are generally conceptualized from the insightful
hypothesis made by \citet{pope_1975}, where the anisotropic Reynolds stress is expressed as an algebraic tensor
polynomial. For example, \citet{PhysRevFluids.3.074602} built a framework for enhancing RANS turbulence models with
carefully designed input features to ensure Galilean and rotational invariance of the model predictions. Following this
work, \citet{doi:10.1063/5.0022561} then improved the approach by reconsidering the criterion for selecting input
features. Another school of thought is to recast the neural network architectures. The pioneering effort could trace
back to the work by \citet{ling_kurzawski_templeton_2016}, in which they proposed a novel deep neural network
architecture that integrated a set of tensor basis into neural networks to model anisotropic Reynolds stress tensor.
Following this effort, \citet{doi:10.1063/5.0048909} recently presented a ML-based turbulence modeling framework in
which two parallel ML modules are combined to directly infer structural and parametric representations of turbulence
physics, respectively.

Despite the improved performance presented in the above-mentioned works and many others \citep{LIU2021108822,
CAO2023123622}, the black-box nature of most ML methods hampers the understanding of obtained data-driven models, making
it difficult to provide physical interpretations and infer new flow physics. On the other hand, it may also increase the
difficulty for disseminating the learned models to end users since there are no explicit mathematical equations.

Recently, symbolic regression approaches have gained a renaissance in ML community. The task of symbolic regression is
to find a symbolic function that best predicts the target given input variables. Some researchers have applied these
approaches to detect hidden fluid properties, and the relevant results present towering promise and preliminary success
\citep{doi:10.1063/1.5136351, doi:10.1063/5.0096669}. Driven by such prevalence, various symbolic regression methods,
such as gene expression programming (GEP) and sparse regression, have been introduced to derive new Reynolds stress
models that can be expressed in explicit algebraic forms. \citet{WEATHERITT201622} proposed an expansion of the GEP
method to model Reynolds stress tensor from high-fidelity data. This work was then further developed by
\citet{ZHAO2020109413} by integrating the CFD simulations with GEP training process. The learning target was set as the
velocities rather than the Reynolds stress itself. However, due to the GEP being by nature a non-deterministic method,
the mathematical form of the model it discovers can vary from different runs. By contrast,
\citet{schmelzer2020discovery} introduced a deterministic symbolic regression method to infer algebraic stress models by
leveraging the fast function extraction method. This method can identify the relevant candidate functions by imposing a
sparsity constraint. Similarly, the sparse identification of non-linear dynamics approach was employed by
\citet{PhysRevFluids.5.084611} to formulate new turbulence closures with affordable interpretability and
transportability.

The success of symbolic regression methods in discovering turbulence models motivates the present work, in which a new
model discovery strategy based on the deep symbolic regression (DSR) method \citep{petersen2020deep} is introduced to
formulate Reynolds stress models that can be explicitly expressed in algebraic forms. The DSR method is a deterministic
methodology and is built on the combination of deep learning and symbolic regression techniques. During the model
training process, the DSR approach can ensure that all samples adhere to all constraints, without rejecting samples post
hoc. In contrast, the GEP method may produce operations that violate imposed constraints, thus requiring rejections post
hoc. This process could be problematic. In addition, most symbolic regression methods suffer from the expectation
problem. That is, these methods are fundamentally designed to optimize the expected performance. However, symbolic
regression generally aims to maximize the performance of few (or single) best-fitting samples. This problem can be
solved by leveraging the risk-seeking reinforcement learning (RL) algorithm, as is done in the present work.

In a sense, the utilization of RL could be one of the most appealing features of DSR method. It is a natural choice
since symbolic space can be considered as an environment where states and actions are given as symbolic tokens
\citep{poesia2021contrastive}. Unlike supervised learning which needs a knowledgeable external supervisor to provide
labeled data, RL is designed for decision-making problems. It generally does not need data to be labeled and can learn
through trial and error \citep{Sutton2018}. RL has gained unprecedented interest in many domains, such as robotics
\citep{doi:10.1146/annurev-control-042920-020211}, mathematics \citep{fawzi2022discovering}, and games
\citep{silver2017mastering}. Recent years also have witnessed a blossoming of RL in fluid mechanics
\citep{doi:10.1063/5.0128446}. Some typical applications include the behavior adaption of swimmers
\citep{PhysRevE.105.045105, PhysRevLett.118.158004}, active flow control for drag reduction \citep{Rabault2019a,
Rabault2019b, doi:10.1063/5.0006492, doi:10.1063/5.0080922} and conjugate heat transfer \citep{HACHEM2021110317,
doi:10.1080/14685248.2020.1797059}, and aerodynamic shape optimization \citep{doi:10.2514/1.J060189,
VIQUERAT2021110080}.

The application of RL in formulating turbulence models is at a very early stage, with most works focusing on the
development of reliable subgrid-scale models for LES. The pioneering study is proposed by \citet{novati2021automating},
in which RL is leveraged to adjust the coefficients of the eddy-viscosity closure model, in order to reproduce the
energy spectrum of homogeneous isotropic turbulence (HIT) predicted by DNS. This approach was then further developed by
\citet{doi:10.1063/5.0106940} for wall-bounded turbulence. The main advantage of RL against supervised learning is that
RL could address the distinction between a priori and a posteriori evaluation and account for compounding modeling
errors \citep{novati2021automating}. In a similar but conceptually different study, \citet{bae2022scientific} applied RL
for the discovery of wall models for LES. Moreover, supervised learning could be ill-posed for turbulence modeling in
implicit filtered LES due to labeled filter form is not available. This challenge can be alleviated by RL, as is done by
\citet{KURZ2023109094}, who applied RL with convolutional neural networks to find an optimal eddy-viscosity for
implicitly filtered LES of HIT.

By integrating the invariants and tensor basis in the DSR model training method, the objective of this work is to
develop a new method for discovering turbulence closures in explicit algebraic forms, with emphases on
\begin{enumerate}
\item Interpretability: the obtained model can be explicitly expressed as a finite tensor polynomial, making it easier
to infer the underlying physics and reveal the constitutive relationship between input features and output targets.
\item Galilean and rotational invariance: by constructing the feature space and the neural network architecture with an
integrity tensor basis and invariants, both Galilean and rotational invariance can be achieved.
\item Transportability: it is straightforward to implement the learned model in RANS solvers since it is expressed as
mathematical expressions. Consequently, there is no need to redeploy the production systems.
\end{enumerate}

The remainder of this paper is organized as follows. First, a brief introduction to the main governing equations used
for performing simulation is provided in Section~\ref{sec:Methodology}, along with the general theory and implementation
details underlying the construction of training methods. Next, three canonical flows with notably different geometry are
selected as testing cases. And the simulation results that highlight the robustness and generalization ability of the
proposed approach are detailed in Section~\ref{sec:results}. Finally, a summary of the contribution is demonstrated in
Section~\ref{sec:conclusions}.

\section{Methodology}\label{sec:Methodology}

\subsection{Governing equations}
In this work, the incompressible turbulent flow is taken to illustrate the modeling framework. The governing RANS
equations read
\begin{equation}
  \nabla \cdot \pmb{u}=0,
\end{equation}
\begin{equation}
  \frac{\partial \pmb{u}}{\partial t} + \pmb{u} \cdot \nabla \pmb{u} = - \nabla p + \nu \nabla^{2} \pmb{u} - \nabla \cdot \pmb{\tau}, \label{eq_mom}
\end{equation}
where $\pmb{u}$, $p$, and $\nu$ are the mean velocity, mean pressure (normalized by density), and viscosity,
respectively. $\pmb{\tau}$ is the Reynolds stress accounting for unresolved turbulence. Herein the Reynolds stress is
unknown and needs to be closed by a mathematical model. Hence, the effects regarding the discretization schemes and
numerical algorithms aside, a turbulence closure model acts as the most critical factor in the prediction accuracy of
RANS equations. In the present work, a symbolic regression method based on deep learning techniques will be leveraged to
produce extra anisotropic terms for Reynolds stress, to improve the predictive accuracy of RANS equations.

\subsection{Baseline RANS turbulence model}

The most commonly used theory for the closure of Reynolds stress tensor is the linear eddy viscosity model (LEVM)
proposed by Boussinesq
\begin{equation}
  \pmb{\tau}=\frac{2}{3} k\pmb{I} - \nu_{t}\left[\nabla\pmb{u}+\left(\nabla\pmb{u}\right)^{\operatorname{T}}\right],
  \label{levm}
\end{equation}
where $\pmb{I}$ denotes the identity matrix, $(\cdot)^{\operatorname{T}}$ means transposition. $\nu_{t}$ indicates the
turbulent viscosity which is unknown and needs to be estimated by one RANS turbulence model. This assumption is widely
used at the expense of accuracy and is susceptible to influences from certain flow configurations, as will be discussed
in the following sections.

The data-driven turbulence model framework developed in the present study is based on the widely applied standard
$k-\varepsilon$ model \citep{launder1974application}, in which the equations for the turbulent kinetic energy $k$ and
the turbulent dissipation rate $\varepsilon$ are written as
\begin{equation}
  \frac{\partial k}{\partial t}+\pmb{u}\cdot\nabla k=\nabla \cdot\left(\nu_{eff, k} \nabla k\right)+P_{k}- \varepsilon, \label{equ:k}
\end{equation}
\begin{equation}
  \frac{\partial\varepsilon}{\partial t}+\pmb{u}\cdot\nabla \varepsilon=\nabla \cdot\left(\nu_{eff, \varepsilon}\nabla \varepsilon\right)+C_{\varepsilon 1} \frac{\varepsilon}{k} P_{k}-C_{\varepsilon 2} \frac{\varepsilon^{2}}{k}, \label{equ:epsilon}
\end{equation}
where other relations for closing the transport equations are
\begin{equation}
  \nu_{eff, k} = \nu+\frac{\nu_{t}}{\sigma_{k}}, \quad \nu_{eff, \varepsilon} = \nu+\frac{\nu_{t}}{\sigma_{\varepsilon}},\quad P_{k} = -\pmb{\tau}: \nabla\pmb{u}.
\end{equation}
Starting with the Boussinesq eddy viscosity assumption, the turbulent eddy viscosity $\nu_{t}$ is given as
\begin{equation}
  \nu_{t}=\frac{C_{\mu} k^{2}}{\varepsilon}.\label{equ:nut}
\end{equation}

Furthermore, the five model coefficients $C_{\mu}, C_{\varepsilon 1}, C_{\varepsilon 2}, \sigma_{k},
\sigma_{\varepsilon}$ are flow-specific tuning and fudge parameters. In the context of this work, the standard values
due to \citet{launder1974application} are adopted. The whole set of these coefficients is shown below
\begin{equation}
  C_{\mu}=0.09, C_{\varepsilon 1}=1.44, C_{\varepsilon 2}=1.92, \sigma_{k}=1.0, \sigma_{\varepsilon}=1.3. \label{kepsilonConstants}
\end{equation}

It is worth stressing that these constant coefficients are calibrated by simplified theories and limited experiments. In
addition, the standard $\varepsilon$ equation (Eq.~\ref{equ:epsilon}) is not based on the exact transport equation. It
is best viewed as an empirical equation since only large-scale flow motions are considered in its derivation. As a
consequence, the accuracy of standard $k-\varepsilon$ model is possibly impeded in turbulent separated flow simulations,
which can be well solved by the data-driven turbulence model developed in this work.

\subsection{Data-driven RANS turbulence framework}

Although the Boussinesq assumption has been widely used in turbulence modeling research, it cannot produce an adequate
approximation to Reynolds stress. In addition, it could become invalid in flow simulations featuring large
separations \citep{doi:10.2514/1.J059953, wilcox1998turbulence}. The goal of this work is to introduce a data-driven
turbulence model with explicit mathematical expressions to increase the accuracy of traditional RANS simulations. Rather
than simply correcting the parameters of traditional turbulence models, this work seeks a more comprehensive effort to
improve the Boussinesq assumption by adding an extra stress tensor $\pmb{b}^{\bot}$ as
\begin{equation}
  \boldsymbol{\tau}=\frac{2}{3} k\pmb{I} - \nu_{t}\left[\boldsymbol{\nabla}\boldsymbol{u}+\left(\boldsymbol{\nabla}\boldsymbol{u}\right)^{\operatorname{T}}\right] + k\pmb{b}^{\bot}.
\end{equation}

\begin{figure*}
  \centering
  \includegraphics[width=0.92\textwidth]{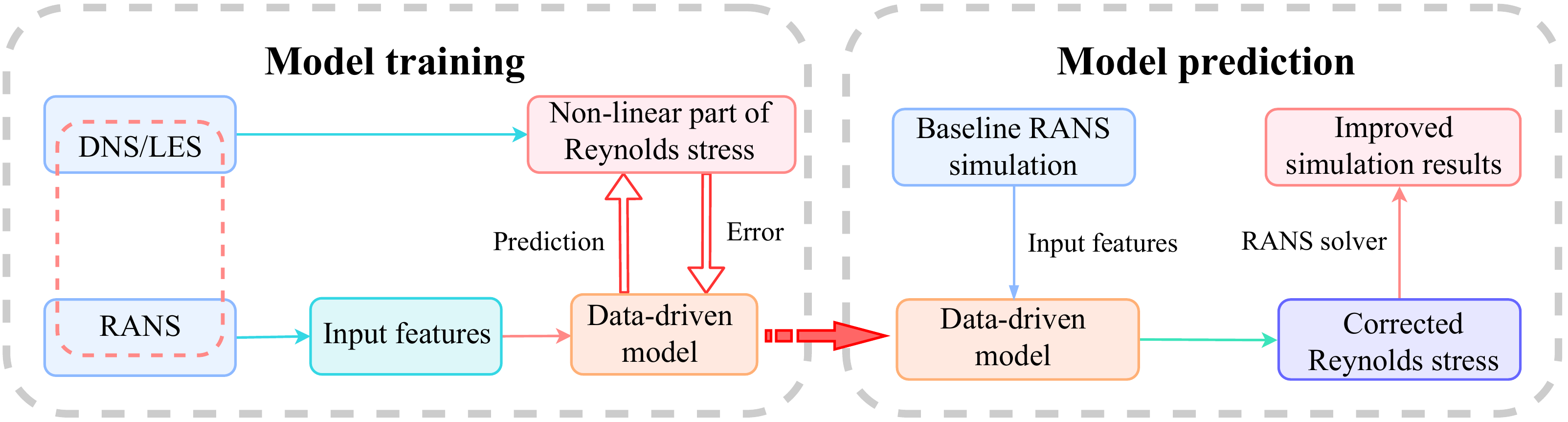}
  \caption{A schematic of the framework for data-driven turbulence modeling. The overall process includes two stages:
  training (left block) and prediction (right block).}
  \label{fig:training_framework}
\end{figure*}

The $\pmb{b}^{\bot}$ represents the non-linear part of Reynolds stress tensor (detailed in next section). Taking
$\pmb{b}^{\bot}$ as the learning target, the framework used in this work is illustrated in
Fig.~\ref{fig:training_framework}. This framework consists of two key phases. During the training phase, a number of
RANS and DNS/LES flow pairs are utilized to build a regression model. The training process aims to minimize the
discrepancy (the training error in Fig.~\ref{fig:training_framework}) between ML predictions and high-fidelity data.
Then, the obtained ML model is used to predict the target flow fields for new unseen flows. This process corresponds to
the prediction phase. The corrected fields are subsequently propagated through the modified RANS solver to improve the
baseline RANS predictions. To avoid the divergence of the data-driven simulations, a common practice is to freeze the
predictions of the ML model. More specifically, the ML model is only called at the initial time and its predictions are
then kept as constants when solving the RANS equations. This mode is usually referred to as loose coupling and is used
in present work. On the contrary, tight coupling requires that the ML model participates in the iterative process of
RANS simulations. And readers who are interested in this subject can refer to the review by
\citet{PhysRevFluids.6.050504}.

The framework shown in Fig.~\ref{fig:training_framework} has permeated the area of data-driven turbulence modeling
research, with various learning targets used by different researchers. In the context of this work, the symbolic
regression method based on deep learning is selected to serve as the ML model. The details about the construction of DSR
method and turbulence modeling schemes will be outlined in the following sections.

\subsubsection{Symbolic regression based on deep learning}\label{sec:dso}

The deep neural network could be the most commonly used regression model for building data-driven models. An immediate
difficulty relates to its interpretability, which tends to be prohibitively tricky since deep learning models are highly
recursive \citep{rudin2019stop}. By contrast, taking compact symbolic formulations to describe physical systems can
provide inherently interpretable insights, thus making it easier to infer with existing theories. In this study, based
on the DSR approach initially proposed by \citet{petersen2020deep}, a data-driven framework is developed for
understanding the mathematical relationships among variables in a turbulent flow system.

\begin{figure*}
  \centering
  \includegraphics[width=0.8\textwidth]{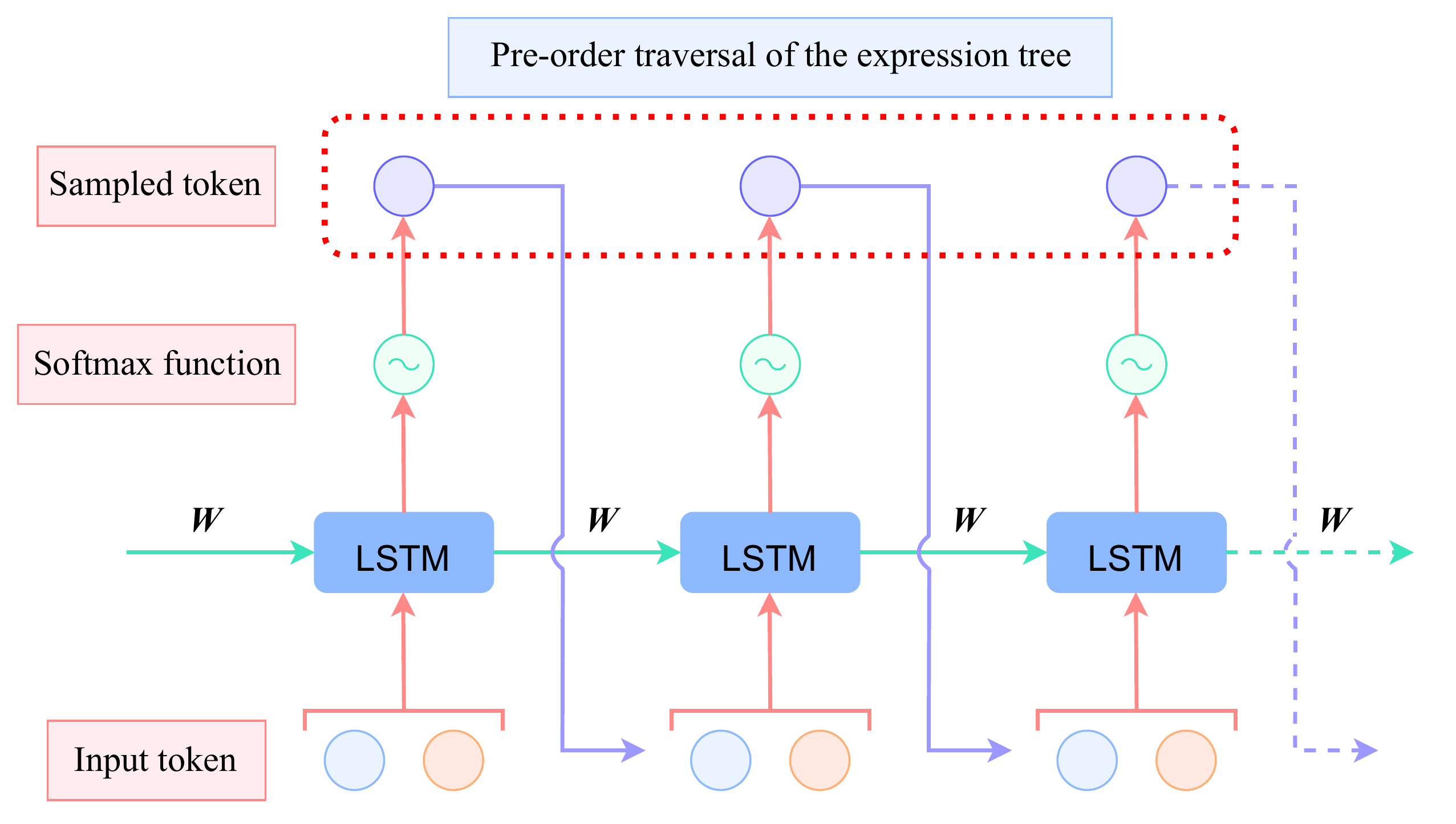}
  \caption{An overview for generating expressions with DSR methodology.}
  \label{fig:sampleToken}
\end{figure*}

DSR is a gradient-based approach for symbolic regression. The core idea of DSR is to use a large model to search the
space of a small model. More specifically, DSR employs expressive recurrent neural networks (RNNs) to generate
best-fitting symbolic expressions under some pre-specified constraints. In DSR, each token for constructing symbolic
expressions is sampled from the outputs of a special kind of RNNs, the long short-term memory (LSTM) neural network. The
fitness of symbolic expressions is then used as the reward function to train the neural networks by a novel RL
algorithm. By this means, it is possible to seamlessly combine the representational capacity of deep learning models and
the natural interpretability of symbolic expressions. An overview of DSR approach is illustrated in
Fig.~\ref{fig:sampleToken}, and the whole training process can be broadly summarized as follows

\begin{enumerate}
  \item Design inputs to LSTM neural networks. LSTM neural networks can learn the dependence among data sequences, which
  means the obtained token would be partially determined by previously sampled tokens. In this sense, the search space
  for symbolic expressions is inherently hierarchical. To ensure the hierarchical information is well-captured, DSR
  leverages the parent and sibling nodes of the token being sampled as inputs to the LSTM networks. Consequently, the
  sampled token is mainly determined by its adjacent nodes in the expression tree.
  \item Impose a priori constraints to the search space. Under the DSR framework, several domain-agnostic constraints
  are applied to reduce the search space
  \begin{itemize}
  \item The minimum and maximum length of symbolic expressions have to be specified before training starts.
  \item A binary operator at least has one child that is non-constant.
  \item The child of a unary operator should not be the inverse of that operator.
  \item The descendants of trigonometric operators cannot contain trigonometric operators.
  \end{itemize}
  These constraints are concurrently applied in the training process by ditching the tokens that would violate any one
  of the constraints. Such a sampling process craftily respects all pre-specified limits without rejecting samples post
  hoc, which makes the DSR method more amenable to complex tasks.
  \item Choose the reward function. Within the DSR approach, the neural networks are trained by a risk-seeking policy
  gradient algorithm, and the performance of sampled expressions is evaluated with a reward function. The symbolic
  expressions that best fit the training dataset could vary with different reward functions. This feature will be
  discussed in detail in the next section.
  \item Optimize constants in expressions. The sampled expressions may include constants that need to be optimized to
  maximize the reward function. In DSR, an inner optimization loop for each sampled expression is performed before
  executing each training epoch.
\end{enumerate}

As can be observed in Fig.~\ref{fig:sampleToken}, a softmax activation function is used in the output layer.
Therefore, the LSTM neural networks would emit a probabilistic distribution over symbolic expressions. The token with
the highest probability is then sampled to constitute the mathematical expressions. If the token being sampled is
against the pre-defined constraints, however, its sampling probability would be reset to zero. After finishing the
training process, the pre-order traversal of sampled tokens can represent the symbolic expression tree in which internal
nodes are mathematical operators and terminal nodes are input variables or constants. The mathematical expression can
then be generated by the pre-order traversal of its corresponding expression tree.

As aforementioned, within the DSR framework, a novel RL algorithm, known as the risk-seeking policy gradient algorithm,
is chosen to train the neural network model. The objective of symbolic regression is to maximize the best-case
performance. This objective could be plagued by traditional policy gradient methods since they are fundamentally
designed to optimize the expected performance (the performance is evaluated by a group of samples). The risk-seeking
policy gradient formulation aims to increase the reward of the top $\epsilon$ fraction of samples from the distribution,
without regard for samples below that threshold. Therefore, this algorithm overcomes a performance bottleneck
encountered in traditional policy gradient methods, enhancing its practical applications.

The update procedures to obtain best-fitted symbolic expressions can be summarized as shown in Algorithm 1. A brief
introduction about the LSTM neural network and the RL algorithm is provided in appendixes~\ref{appendix:lstm} and
\ref{appendix:rspg}. For complete details about the DSR algorithm, the reader is invited to consult the original works
by \citet{petersen2020deep}.
\begin{table*}[]
  \centering
  \begin{tabular}{@{}lr@{}}
  \toprule[2pt]
  \multicolumn{2}{l}{\textbf{Algorithm 1}: Symbolic regression based on reinforcement learning} \\ \midrule[1pt]
  \multicolumn{2}{l}{\textbf{Input}: learning rate $\alpha$; entropy coefficient $\lambda_{\mathcal{H}}$; risk factor $\epsilon$; batch size $N$; reward function $R$.}\\
  \multicolumn{2}{l}{\textbf{Goal}: A symbolic mathematical expression that can best fit the dataset.} \\
  \multicolumn{2}{l}{\textbf{Initialize}: LSTM neural network with parameters $\theta$, defining distribution over expression $p(\cdot \mid \theta)$.} \\
  \multicolumn{2}{l}{\textbf{For} each epoch \textbf{do}} \\
  \quad 1: $\mathcal{T} \leftarrow\left\{\tau^{(i)} \sim p(\cdot \mid \theta)\right\}_{i=1}^N$ & Sample $N$ expressions\\
  \quad 2: $\mathcal{T} \leftarrow\left\{OptimizeConstants(\tau^{(i)}, R)\right\}_{i=1}^N$ & Optimize constants w.r.t. reward function \\
  \quad 3: $\mathcal{R} \leftarrow\left\{R(\tau^{(i)})\right\}_{i=1}^N$ & Compute rewards \\
  \quad 4: $\mathcal{R}_{\epsilon} \leftarrow\left\{(1-\epsilon)- quantile\; of \; \mathcal{R}\right\}$ & Compute reward threshold \\
  \quad 5: $\mathcal{T} \leftarrow\left\{\tau^{(i)}: R(\tau^{(i)})\geq R_{\epsilon}\right\}$ & Select subset of expressions above threshold\\
  \quad 6: $\mathcal{R} \leftarrow\left\{R(\tau^{(i)}): R(\tau^{(i)})\geq R_{\epsilon}\right\}$ & Select corresponding subset of rewards \\
  \quad 7: $\hat{g}_{1} \leftarrow ReduceMean((\mathcal{R}-\mathcal{R}_{\epsilon})\nabla_{\theta}\log p(\mathcal{T} \mid \theta))$ & Compute risk-seeking policy gradient \\
  \quad 8: $\hat{g}_{2} \leftarrow ReduceMean(\lambda_{\mathcal{H}}\nabla_{\theta}\mathcal{H}(\mathcal{T}\mid\theta))$ & Compute entropy gradient \\
  \quad 9: $\theta \leftarrow \theta + \alpha(\hat{g}_{1} + \hat{g}_{2})$ & Apply gradients \\
  \quad 10: \textbf{If} max $\mathcal{R} > R(\tau^{*})$ \textbf{then} $\tau^{*} \leftarrow \tau^{(arg\;max\;\mathcal{R})}$ & Update best expression \\
  \multicolumn{2}{l}{\textbf{End}}\\
  \multicolumn{2}{l}{\textbf{Return} $\tau^{*}$} \\
  \bottomrule[2pt]
  \end{tabular}
\end{table*}

\subsubsection{DSR for turbulence model development}

Rather than recovering exact physical models from data, in the context of this work, the DSR method is employed to
identify turbulence closures by searching over the space of tractable mathematical expressions to best fit the dataset
from high-fidelity flow simulations. The theoretical foundation for this framework is the stress tensor decomposition
proposed by \citet{pope_1975}. Taking the turbulent kinetic energy $k$ and turbulence dissipation rate $\varepsilon$ as
two scaling parameters, the Reynolds stresses and the rates of stain can be normalized as follows
\begin{equation}
  \pmb{b} = \pmb{\tau}/k-\frac{2}{3}\pmb{I},
\end{equation}
\begin{equation}
  \pmb{S}=\frac{1}{2} \frac{k}{\varepsilon}\left[\nabla\pmb{u}+\left(\nabla\pmb{u}\right)^{\operatorname{T}}\right], \quad \pmb{R}=\frac{1}{2} \frac{k}{\varepsilon}\left[\nabla\pmb{u}-\left(\nabla\pmb{u}\right)^{\operatorname{T}}\right],
\end{equation}
where $\pmb{I}$ is identity matrix, $(\cdot)^{\operatorname{T}}$ denotes matrix transposition.

It is postulated that $\pmb{S}$ and $\pmb{R}$ contain all information for determining $\pmb{b}$ and any tensor can be
expressed by an infinite tensor polynomial. Leveraging the Cayley-Hamilton theorem, Pope proved that the normalized
Reynolds stress tensor can be expressed as a finite polynomial linearly composed of tensor bases and scaler
functions
\begin{equation}
  \pmb{b}(\pmb{S}, \pmb{R})=\sum_{m} G^{m}\left(I_{1}, \ldots, I_{n}\right) \pmb{T}^{m}, \label{equ:pope_b}
\end{equation}
where $G^{m}$ are several scalar coefficients determined by the invariants $I_{n}$, and $\pmb{T}^{m}$ are
independent, symmetric tensor basis functions. For statistically two-dimensional flows, as is the case in present work,
the coefficients are merely dependent on at most two invariants (i.e., $n=2$) and the bases can be simplified to only
three tensors (i.e., $m=3$), as shown below
\begin{equation}
  \begin{array}{lll}
    \pmb{T}^{1}=\pmb{S}, & \pmb{T}^{2}=\pmb{S R}-\pmb{R S}, & \pmb{T}^{3}=\pmb{S}^{2}-\frac{1}{3} \pmb{I} \cdot \operatorname{Tr}\left(\pmb{S}^{2}\right),
  \end{array}\label{equ:tensor_basis}
  \end{equation}
\begin{equation}
  I_{1}=\operatorname{Tr}\left(\pmb{S}^{2}\right), \quad I_{2}=\operatorname{Tr}\left(\pmb{R}^{2}\right),\label{equ:inv}
\end{equation}
where $\operatorname{Tr\left(\cdot\right)}$ denotes the trace of matrix.

\begin{figure*}
  \centering
  \includegraphics[width=0.85\textwidth]{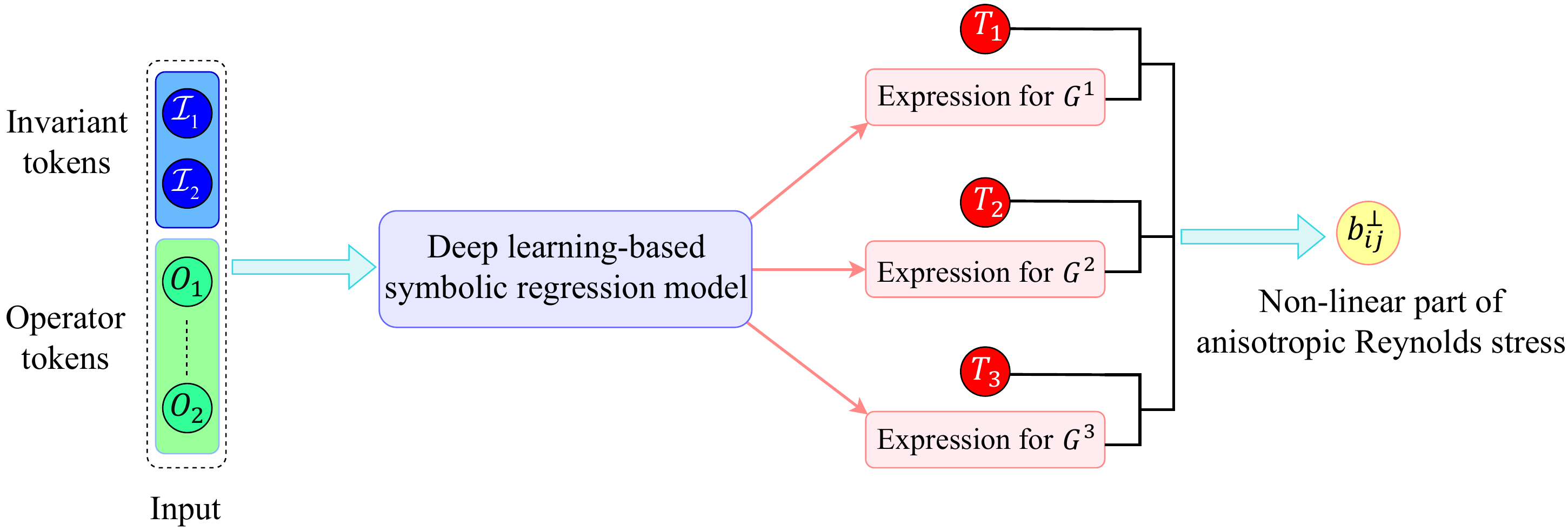}
  \caption{Illustration of the DSR framework utilized in present work for building data-driven turbulence closures.}
  \label{fig:framework}
\end{figure*}

The utilization of invariants and tensor basis functions is the key to keep the Galilean and rotational invariance of
turbulence model \citep{ling_kurzawski_templeton_2016}. More generally, Embedding the invariance into the input features
could better improve the model performance, which has been highlighted by some early works \citep{doi:10.1063/1.5132378,
kutz_2017}. In the present work, the DSR approach is utilized to generate symbolic expressions for the coefficients,
given the invariants and tensor bases. A simplified overview of the DSR framework used in present study is schematically
depicted in Fig.~\ref{fig:framework}. Since the flow simulations in present work are statistically in two dimensions,
the input tokens are composed of two invariants and several other constants and mathematical operators. The three
mathematical expressions produced by DSR correspond to the scalar coefficients $G^1$, $G^2$, and $G^3$, respectively.
The Reynolds stress can then be represented by the linear combination of mathematical expressions and tensor bases. The
discrepancy between the Reynolds stress calculated by symbolic expressions and that from high-fidelity simulations is
taken as the reward signal to inform the training process so as to maximize the performance of best-fitting expressions.

As is indicated in Fig.~\ref{fig:framework}, however, only the non-linear part of Reynolds stress will be built by the
developed framework. More specifically, instead of taking $\pmb{b}$ directly as the modeling subject, the deviatoric
Reynolds stress is split into two portions as expressed by
\begin{equation}
  \begin{aligned}
    \pmb{b} &= \pmb{b}^{\Vert} + \pmb{b}^{\bot} \\
    &= -2C_{\mu}\pmb{S} + \pmb{b}^{\bot},
  \end{aligned}\label{equ:splitRS}
\end{equation}
where $\pmb{b}^{\Vert}$ and $\pmb{b}^{\bot}$ denote the linear and non-linear parts of Reynolds stress, respectively.
Recalling that the standard $k-\varepsilon$ turbulence model is used as the starting point for building data-driven
models, the constant parameter $C_{\mu}$ belonging to $k-\varepsilon$ turbulence model (see Eq.~\ref{kepsilonConstants})
is kept in the linear portion. This linear part will be taken as classical LEVM and solved implicitly in modified RANS
solvers. As a consequence, the framework presented in Fig.~\ref{fig:framework} can be accordingly expressed in
mathematics as
\begin{equation}
  \pmb{b}^{\bot}=\sum_{m=1}^{3} G^{m}\left(I_{1}, I_{2}\right) \pmb{T}^{m}, \label{equ:pre_b}
\end{equation}
where coefficients $G$ will be represented by symbolic expressions consisting of invariants, constants and mathematical
operators. In other words, only the non-linear portion of the Reynolds stress tensor, $\pmb{b}^{\bot}$, will be
predicted by the data-driven turbulence model.

This strategy aligns with the stability and accuracy analysis of data-driven turbulence models. In some earlier works
\citep{ling_kurzawski_templeton_2016, GENEVA2019125}, the predictions from ML models are directly injected into RANS
solvers. The remaining flow quantities are subsequently propagated forward by solving RANS equations with a frozen
Reynolds stress field. Although prior assessments show that the discrepancy between ML predictions and high-fidelity
data is pretty small, other derived flow variables of ultimate interest, such as velocity and pressure, are possibly far
removed from the true values for flows with higher Reynolds numbers. Recently, a few studies
\citep{wu_xiao_sun_wang_2019, brener_cruz_thompson_anjos_2021} showed that embedding ML prediction, i.e. the Reynolds
stress field, explicitly into RANS equations would inevitably result in the lack of accuracy of data-driven turbulence
models. The minor error in predicted Reynolds stress would be significantly magnified during the forward iterations of
RANS equations, thus leading to severe performance degeneration of data-driven RANS simulations. To avoid being
ill-conditioned, it is necessary to decompose the Reynolds stress before it is injected into RANS simulations, as is
done herein. This strategy has been proven to be effective in improving the stability and accuracy of data-driven RANS
simulations \citep{wu_xiao_sun_wang_2019, brener_cruz_thompson_anjos_2021, PhysRevFluids.5.084611}.

% \subsection{Implementation for training}

\subsection{Numerical setup}

All simulations in the present works are taken by the open-source CFD platform OpenFOAM \citep{doi:10.1063/1.168744}.
The data-driven simulations can be divided into two rounds. In the first round, the baseline RANS simulation is
performed by using the semi-implicit method for pressure-linked equations (SIMPLE) algorithm to achieve a converged
solution. In the second round, a modified solver based on SIMPLE algorithm accepts the newly corrected non-linear part
of the anisotropic Reynolds stress field. These corrections are then injected into RANS equations and iteratively solved
until the residual reconverges. It should be emphasized that the turbulence closure consisting of symbolic expressions
is only called once before the second round simulation starts, which means the predicted $\pmb{b}^{\bot}$ (non-linear
part of anisotropic Reynolds stress tensor) is kept constant during the iteration of RANS equations.

For discretizing the RANS equations, the second-order central difference scheme is chosen for all terms except for the
convection term, for which the second-order upwind scheme is selected. For all flow cases, the mesh independence
experiment is first carried out before iteration starts, and all meshes are non-uniform with increased resolution around
the feature of interest.

\subsection{Summary of proposed approach}
The entire procedures for implementing the data-driven turbulence model framework based on deep learning-based symbolic
regression can be summarized as follows
\begin{enumerate}
  \item Obtain the training dataset consisting of baseline RANS and DNS simulation data pairs. And run baseline RANS
  simulations for the testing flow set using the standard $k-\varepsilon$ turbulence model (the baseline RANS solution
  will serve as the initial state for second round simulation evolved with symbolic predictions).
  \item Obtain the input features from the training flow set, i.e., the invariants $I_{n}$ and tensor basis
  functions $\pmb{T}^{m}$ from baseline RANS solutions.
  \item Interpolate the DNS results to the RANS computational grids and then obtain the non-linear part of anisotropic
  Reynolds stress tensor (training target), i.e., $\pmb{b}^{\bot}$.
  \item Construct the mapping function $g:\omega_{i}(\pmb{S}, \pmb{R})\mapsto\pmb{b}^{\bot}$ by DSR approach. The
  hyperparameters are carefully tuned before they are applied to the training.
  \item Integrate the symbolic expressed with simulation code. Replace the Reynolds stress term in the RANS solver with
  $-2C_{\mu}\pmb{S}+\pmb{b}^{\bot}$, and then run a forward simulation until residual re-converges. As aforementioned,
  the loose coupling strategy (the symbolic predictions are only called once) is adopted in this work. So the non-linear
  part $\pmb{b}_{*}^{\bot}$ is frozen while the linear part $-2C_{\mu}\pmb{S}$ is iteratively updated during the
  data-driven simulation.
\end{enumerate}

\section{Results}\label{sec:results}

In this section, different flow configurations are investigated to verify the method proposed in this work. These flows
are characterized by massive separations, which are difficult for traditional RANS turbulence models to perform accurate
predictions. The corrected Reynolds stress tensor is substituted into the solver to obtain an improved mean flow field.
The computation framework is verified by comparing the reconverged flow field with the high-fidelity mean flow,
presenting a generalization performance on massively varying geometries.

\subsection{Case setup for training and testing dataset}

\begin{figure*}
  \centering
  \subfigure[]{
  \includegraphics[width=0.6\textwidth]{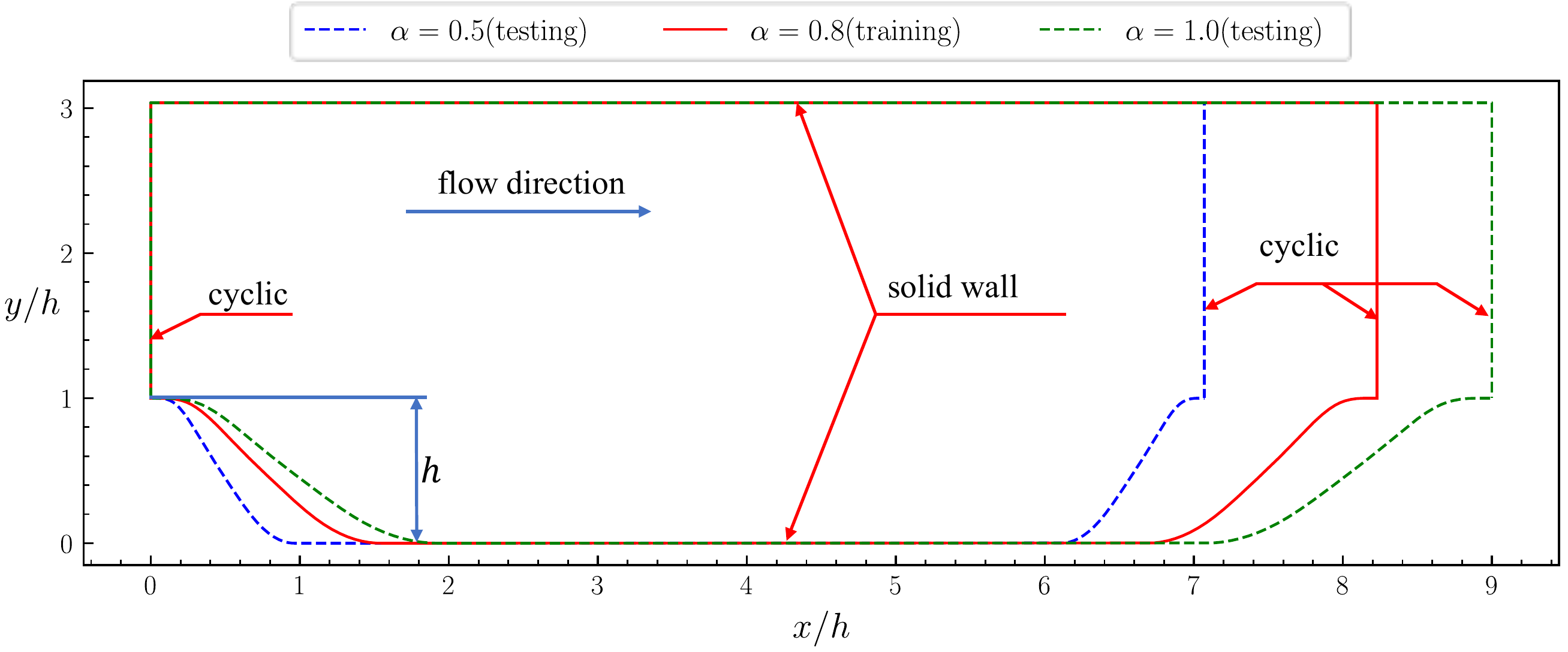}
  }
  \subfigure[]{
  \includegraphics[width=0.8\textwidth]{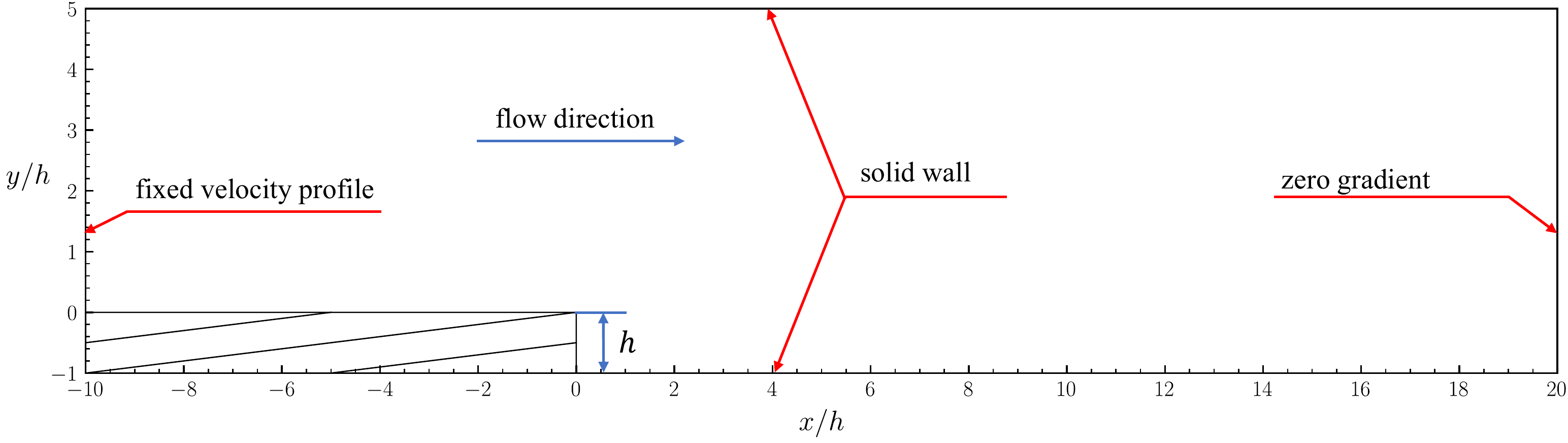}
  }
  \caption{Schematic of the geometries of (a) periodic hills and (b) backward-facing step.}
\label{fig:flowGeo}
\end{figure*}

To assess the performance of the proposed framework, the classical cases of flow over parameterized periodic hills are
selected, as shown in Fig.~\ref{fig:flowGeo} (a). The left and right sides are connected to achieve periodicity. Hence,
the cyclic boundary conditions are used at the inlet (left side) and outlet (right side). The geometries and DNS results
are provided by \citet{XIAO2020104431}. A parameter $\alpha$ is utilized to change the hill steepness and the overall
length of the geometry. The shape parameters for these three cases correspond to $\alpha=0.8$, 0.5, and 1.0,
respectively. The former is used as a training dataset. The obtained symbolic expression for turbulence closure is
further applied to the latter two cases. In addition, the flow through a backward-facing step, which is significantly
different from training data in geometry, is also selected to test the scope of application of the learned model, as
shown in Fig.~\ref{fig:flowGeo} (b). The geometry, DNS results, and experimental data are obtained from the works of
\citet{jovic1994backward} and \citet{le_moin_kim_1997}. For this backward-facing step case, A fixed velocity condition
is enforced at the inlet, and a zero gradient condition is applied for the outflow. The no-slip boundary condition is
applied at the top and bottom walls for all flow cases.

\begin{figure}
  \centering
  \includegraphics[width=0.45\textwidth]{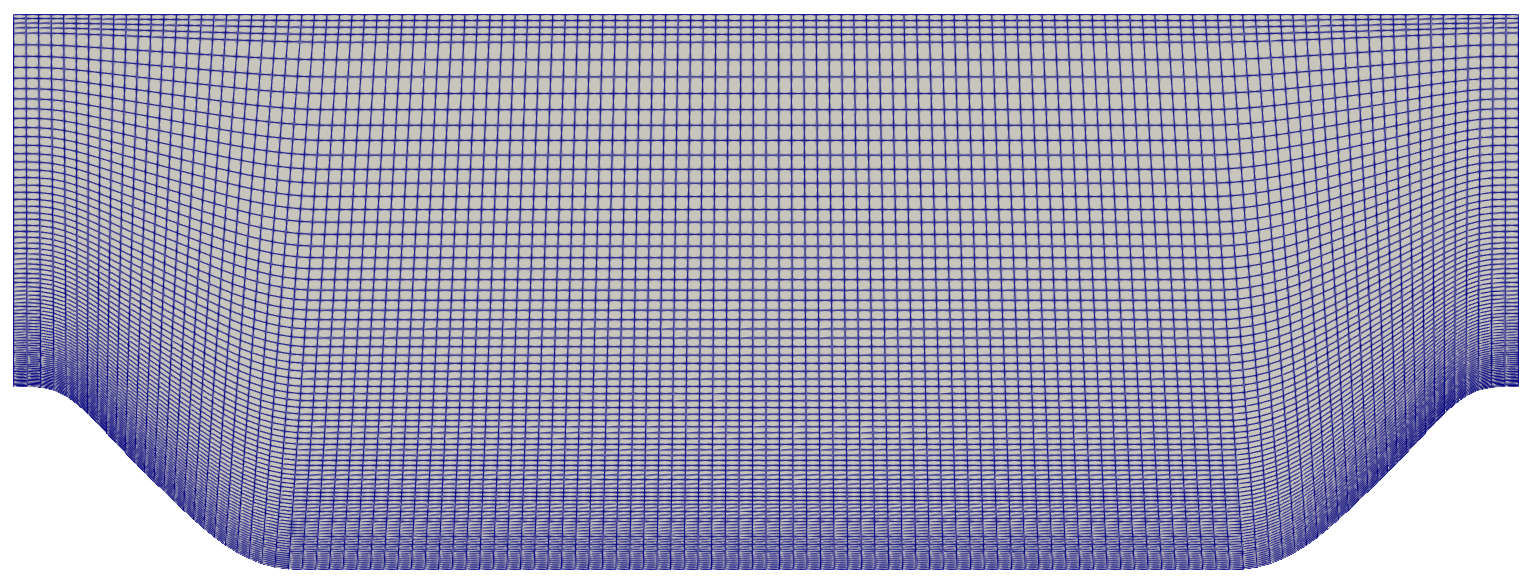}
  \caption{Computation mesh of the periodic hill with $\alpha=0.8$.}
  \label{fig:08Mesh}
\end{figure}

Other information about the training and testing cases, such as Reynolds number and Iterative steps for achieving
convergence (residual of streamwise velocity is around $1\cdot 10^{-10}$), are summarized in Tab.~\ref{tab:train_test}.
A typical mesh description of periodic hill with $\alpha=0.8$ is presented in Fig.~\ref{fig:08Mesh}. The computational
time may vary with hardware, but the simulation should achieve proper convergence performance within hundreds of seconds
on a modern multi-core workstation. Additionally, it is noted that the converge cost with discovered turbulence model is
basically the same with $k-\varepsilon$ computation cost, as has been reported by previous studies
\citep{doi:10.1063/5.0022561}.

\begin{table*}
  \begin{center}
    \caption{Summary of flow cases used for training and testing.}
\def~{\hphantom{0}}
\resizebox{\textwidth}{!}{
  \scriptsize
  \begin{tabular}{cccccc}
    \toprule[2pt]
    Case use        & Description  & Cell number & Reynolds number & Characteristic length $h$ & Iterative steps \\ \midrule[1pt]
    Train & Periodic hill with $\alpha=0.8$ & Around 9600 & 5600 & Hill height & 20000  \\
    Test  & Periodic hill with $\alpha=0.5$ & Around 9600 & 5600 & Hill height & 20000  \\
    Test  & Periodic hill with $\alpha=1.0$ & Around 9600 & 5600 & Hill height & 20000  \\
    Test  & Backward-facing step            & Around 46000 & 5000 & Step height & 20000  \\
    \bottomrule[2pt]
  \end{tabular}}
  \label{tab:train_test}
  \end{center}
\end{table*}

\subsection{Turbulence closure discovery}

The traditional RANS turbulence models share a common weakness in correctly predicting the Reynolds stress anisotropy,
which is blamed for the poor accuracy in many RANS-based flow simulations, especially in those with strong flow
separations. As previously mentioned, a RL algorithm, known as the risk-seeking policy gradient
algorithm, is employed to train the neural network. The non-linear part of anisotropic Reynolds stress tensor is set as
the training target. As a consequence, taking the root-mean-square-error function, the reward function of DSR approach
is accordingly defined as

\begin{equation}
  \frac{1}{1 + \frac{1}{\sigma}\sqrt{\frac{1}{n}\sum\limits_{i=1}^{n}{\|\widehat{\pmb{b}^{\bot}_i} - \pmb{b}^{\bot}_i\|^2}}},\label{equ:reward_rmse}
\end{equation}
where $\widehat{\pmb{b}^{\bot}}$ is the true value of the non-linear part of anisotropic Reynolds stress tensor obtained
from DNS results and $\pmb{b}^{\bot}$ is the corresponding prediction by symbolic expressions. $n$ corresponds to the
size of training dataset. $\sigma$ denotes the standard deviation of $\widehat{\pmb{b}^{\bot}}$. Here, the
root-mean-square-error function is normalized by $\sigma$ and is squashed to bound the range of reward value to (0, 1].

\begin{figure}
  \centering
  \includegraphics[width=0.5\textwidth]{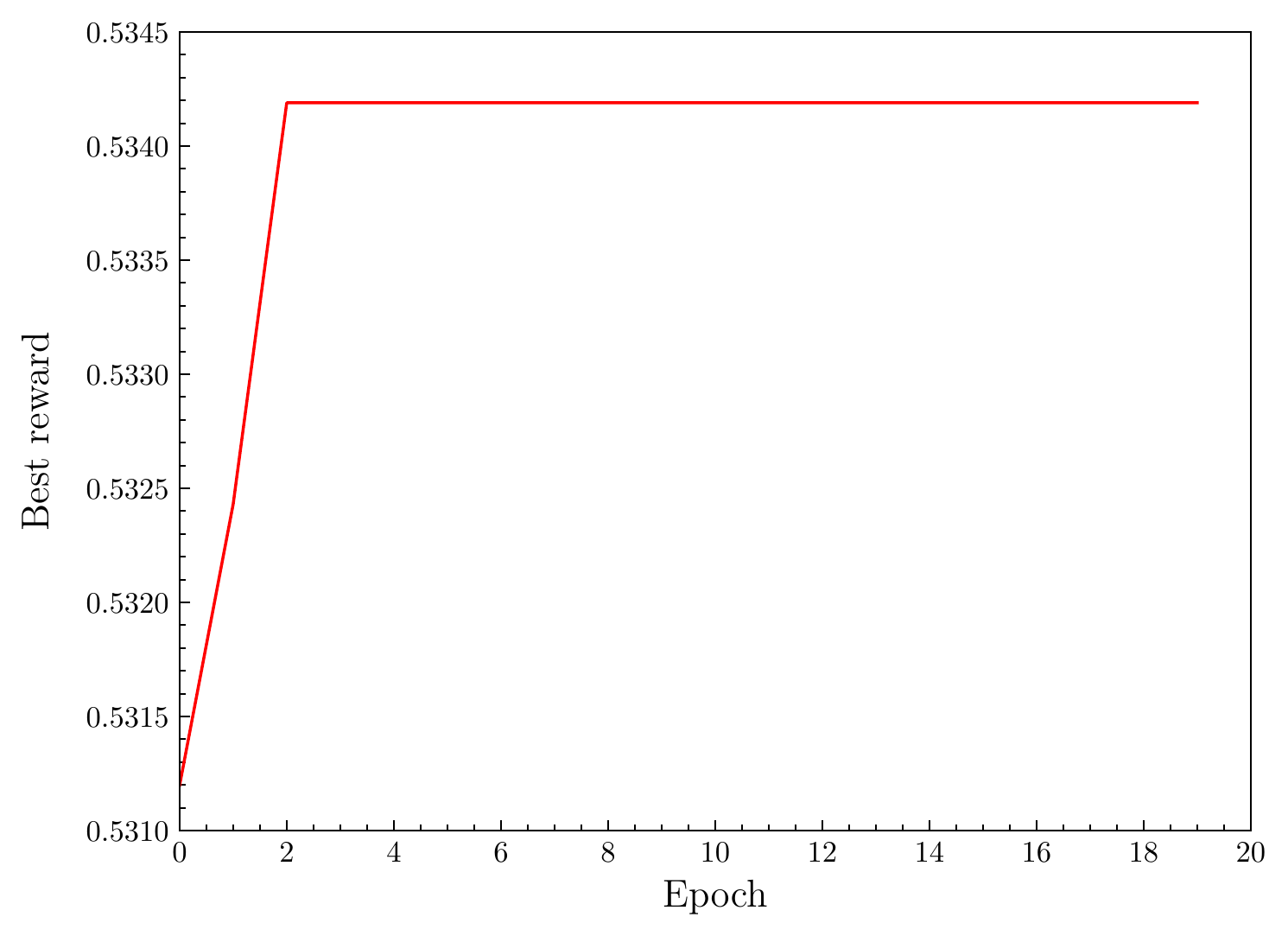}
  \caption{Reward training curve for using DSR to discover turbulence closure models. The curve shows the best reward value as a function of total expressions evaluated so far.}
  \label{fig:reward}
\end{figure}

The main hyper-parameters used for symbolic training are listed in Tab.~\ref{tab:hyperPara}, and the reward curve is
illustrated in Fig.~\ref{fig:reward}, as a function of total expressions evaluated during training. The tokens used to
build symbolic expressions contain four regular arithmetic mathematical operators (+, -, $\times$, $\div$), constants,
and input variables (i.e., the invariants). These tokens are selected to offer a compromise between the readability of
the turbulence closure equation and prediction accuracy. Using non-linear functions such as $\sin$ and $\tan$ could
slightly improve the accuracy of DSR results, but the obtained expressions may be extremely complex. In addition, the
computational cost could also be largely increased. Using the present configurations, the training takes about 100 hours
on a workstation with 20 processor cores (Intel Xeon CPU E5-2650).
\begin{table}[]
  \centering
  \caption{Main DSR hyper-parameters used for searching symbolic expressions of turbulence closures.}
  \label{tab:hyperPara}
  \begin{tabular}{@{}ll@{}}
  \toprule[2pt]
  LSTM architecture (3 hidden layers) & 64 $\rightarrow$ 64 $\rightarrow$ 64 \\
  Train epochs & 20 \\
  Batch size & 640 \\
  Learning rate & 0.001 \\
  Entropy coefficient & 0.005 \\
  Risk factor & 0.05 \\
  Minimum length of expressions & 4 \\
  Maximum length of expressions & 32 \\
  Mathematical operators & [+, -, $\times$, $\div$] \\
  Constant token repeats & 3 \\
  \bottomrule[2pt]
  \end{tabular}
\end{table}

The goal of training efforts is to determine the mathematical expressions of scalar coefficients $G^{m}\left(I_{1},
I_{2}\right)$. To improve the training efficiency, the input invariants to the neural network are re-scaled by a
sigmoidal function due to their strong varying in magnitude
\begin{equation}
  \widehat{I}_{i}=\frac{1-e^{-I_{i}}}{1+e^{-I_{i}}}.
\end{equation}
This operation consequently normalizes the input data to the range [-1, 1], thus reducing the negative impact of some
possible outliers on the model performance. Without causing ambiguity, the symbol $\widehat{(\cdot)}$ of normalized
quantities are briefly dropped in the remainder of this work.

As indicated in Algorithm 1, the constant placeholders in sampled expressions need to be optimized to maximize the
reward function. In this work, a non-linear optimization algorithm, known as Broyden-Fletcher-Goldfarb-Shanno (BFGS)
algorithm \citep{JMLR:v13:fortin12a}, is leveraged to substitute the constant placeholders with optimized constants.
Since running constant optimization can be prohibitively expensive, the number of constant placeholders in each
expression is limited to 3 during training.

The learned model takes the form ($\beta=7$)
\begin{equation}
    \pmb{b}^{\bot} = \frac{\beta}{10}\left(G^{1}\pmb{T}^{1} + G^{2}\pmb{T}^{2} + G^{3}\pmb{T}^{3}\right),\label{equ:bbot}
\end{equation}
where
\begin{equation}
  \left.\begin{array}{l}
    G^{1}=0.1893I_1 + 0.2229I_2 + 0.1176 \\
    G^{2}=-0.1036I_1I_2^3 - 0.05182I_1^2I_2^2 + 0.1718I_1^2 - 0.2333 \\
    G^{3}=-2.514I_1I_2^4 - 3.514I_2^3 - 0.01105I_2^2 - 2I_1I_2 + 2.98I_2
  \end{array}\right\}\label{equ:coefficients}.
\end{equation}
In this learned expression for $\pmb{b}^{\bot}$ (Eq.~\ref{equ:bbot}), the damping factor $\beta/10$ is selected
following the suggestion by \citet{shih1993realizable}. While a detailed discussion can be found therein, this factor is
leveraged to ensure realizability conditions of the Reynolds stress tensor. A similar strategy was also utilized by
\citet{PhysRevFluids.5.084611}. As is indicated by Eq.~\ref{equ:coefficients}, the scalar coefficient expression
multiplying by a higher-order basis tensor exhibits a much more complex form than the expression multiplying by a
low-order basis as the former contains more high-order invariants. This result implies that the higher-order terms could
yield relatively complex qualitative behavior in turbulent flow simulations. In addition, the coefficient expressions
are functions of the invariants herein. By contrast, the turbulence closure models discovered by some other approaches,
such as sparse regression and fast function extraction, only contain constant coefficients that keep unchanged for the
whole computational domain. As a consequence, the method proposed in this work shows a degree of superiority since the
coefficients vary with the invariants at different grid points, which tends to produce a more robust closure model.

\subsection{Predictive results}\label{sec:08Results}

The discovered model is first validated \textit{a prior} for the training periodic hill flow case ($\alpha=0.8$). Within
this effort, the accuracy of the learned model is evaluated by high-fidelity data based on the predicted anisotropic
Reynolds stress tensor. In the following context, the unnormalized anisotropic Reynolds stress tensor $\pmb{a}=k \cdot
\pmb{b}(\pmb{S}, \pmb{R})$ will be referred to as the tensor used for solving RANS equations.

\begin{figure*}
  \centering
  \includegraphics[width=0.98\textwidth]{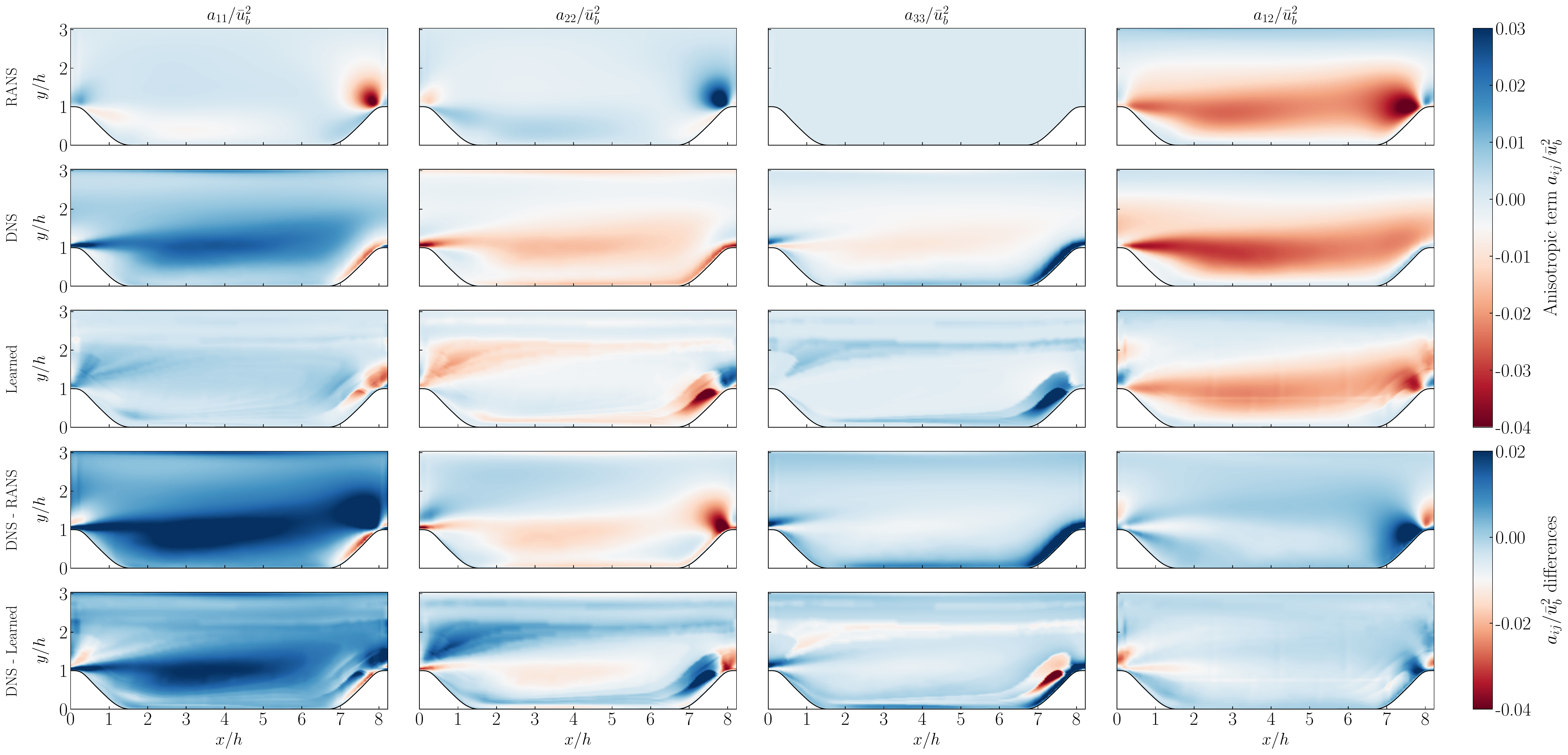}
  \caption{Comparison of anisotropic Reynolds stress components for periodic hill training flow ($\alpha=0.8$) using
  LEVM, DNS, and the learned model, respectively. From left to right: $a_{11}$, $a_{22}$, $a_{33}$, and $a_{12}$. From
  top to bottom: baseline RANS solutions, time-averaged DNS solutions, the learned model solutions, and the difference
  between exact data and predicted data by LEVM and the learned model, respectively. All results are normalized by the
  bulk velocity at crest $\bar{u}_b$.}
  \label{fig:08RS}
\end{figure*}

As is presented in Fig.~\ref{fig:08RS}, the baseline RANS simulation provides a reasonably good prediction for shear
component $a_{12}$, even though some over-predictions can be observed at the downstream crest. However, its predictions
for other three diagonal components are far removed from high-fidelity DNS observations. Such inaccuracy is more obvious
for $a_{22}$ and $a_{33}$, of which the prediction bias is particularly stark near the bottom wall. By contrast, the
learned model captures the correct sign and presents a higher precision in the magnitude for all components of the
anisotropic Reynolds stress tensor. Although the improvement can be plagued at certain flow areas, the discovered model
generally provides better predictions than LEVM. This superiority is relatively more significant for $a_{11}$ and
$a_{22}$ in the region where flow separations could exist. The last two rows in Fig.~\ref{fig:08RS} demonstrate the
discrepancy between the highly-resolved DNS data and the results obtained from LEVM and the learned model, respectively.
It indicates that the learned model is, to some extent, more capable to reduce the predictive error in the anisotropic
stress tensor.

\begin{figure*}
  \centering
  \includegraphics[width=0.98\textwidth]{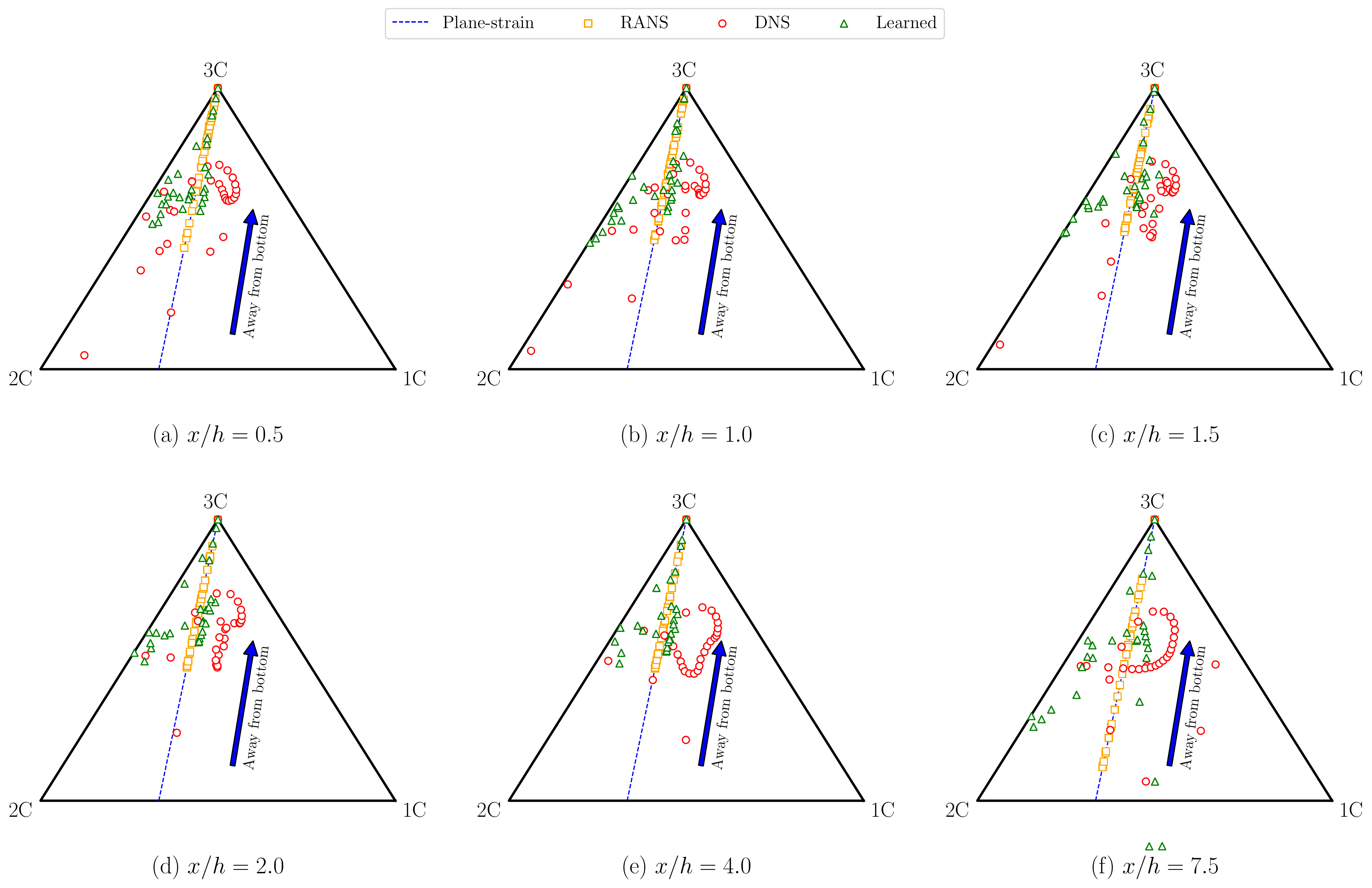}
  \caption{Barycentric map of the predicted Reynolds stress anisotropy for periodic hill training flow ($\alpha=0.8$).
  The learned predictions on six streamwise locations at $x/h=0.5$, $x/h=1.0$, $x/h=1.5$, $x/h=2.0$, $x/h=4.0$ and
  $x/h=7.5$ are compared with the corresponding results from high-fidelity DNS simulations and the standard
  $k-\varepsilon$ RANS turbulence model in (a)-(f), respectively.}
  \label{fig:08tri}
\end{figure*}

As aforementioned, the independent basis tensors and invariants representing key turbulence features are leveraged
during the training process to guarantee both Galilean and rotational invariance of the discovered model. To make a
deeper investigation into the realizability of the discovered turbulence model, the barycentric map
\citep{doi:10.1080/14685240701506896} based on the combination of eigenvalues is used to provide a non-distorted visual
representation of the turbulence anisotropy. The anisotropic Reynolds stress tensor predicted by the discovered symbolic
model is shown in Fig.~\ref{fig:08tri}, compared with baseline RANS and DNS results. The comparisons are performed on
six streamwise locations at $x/h=1.5$, $x/h=1.0$, $x/h=1.5$, $x/h=2.0$, $x/h=4.0$ and $x/h=7.5$, respectively. As
indicated in Fig.~\ref{fig:08tri}, the Reynolds stress anisotropy approaches the three-component limiting state for all
three models as the point moves away from the bottom wall. This trend can ascribe to the significantly fewer flow
separations in the bulk region, thus the turbulence gradually develops to be isotropic. It is also noted that the
baseline RANS Reynolds stress is close to the plane-strain limiting state at all six locations, showing significant
discrepancies from the DNS results. The baseline RANS flow field is generally dominated by the shear layer, thus the
medium eigenvalue of baseline RANS stress anisotropy is close to zero \citep{doi:10.1080/14685240701506896}. By
contrast, the discovered model is more capable to capture the stress anisotropy to achieve better agreement with DNS
data, even though there are a tiny minority of outliers as shown in Fig.~\ref{fig:08tri} (c). It is worth stressing that
these few outliers could hardly impede the performance of the learned model, as is discussed in the following context.

\begin{figure*}
  \centering
  \subfigure[]{
  \includegraphics[width=0.45\textwidth]{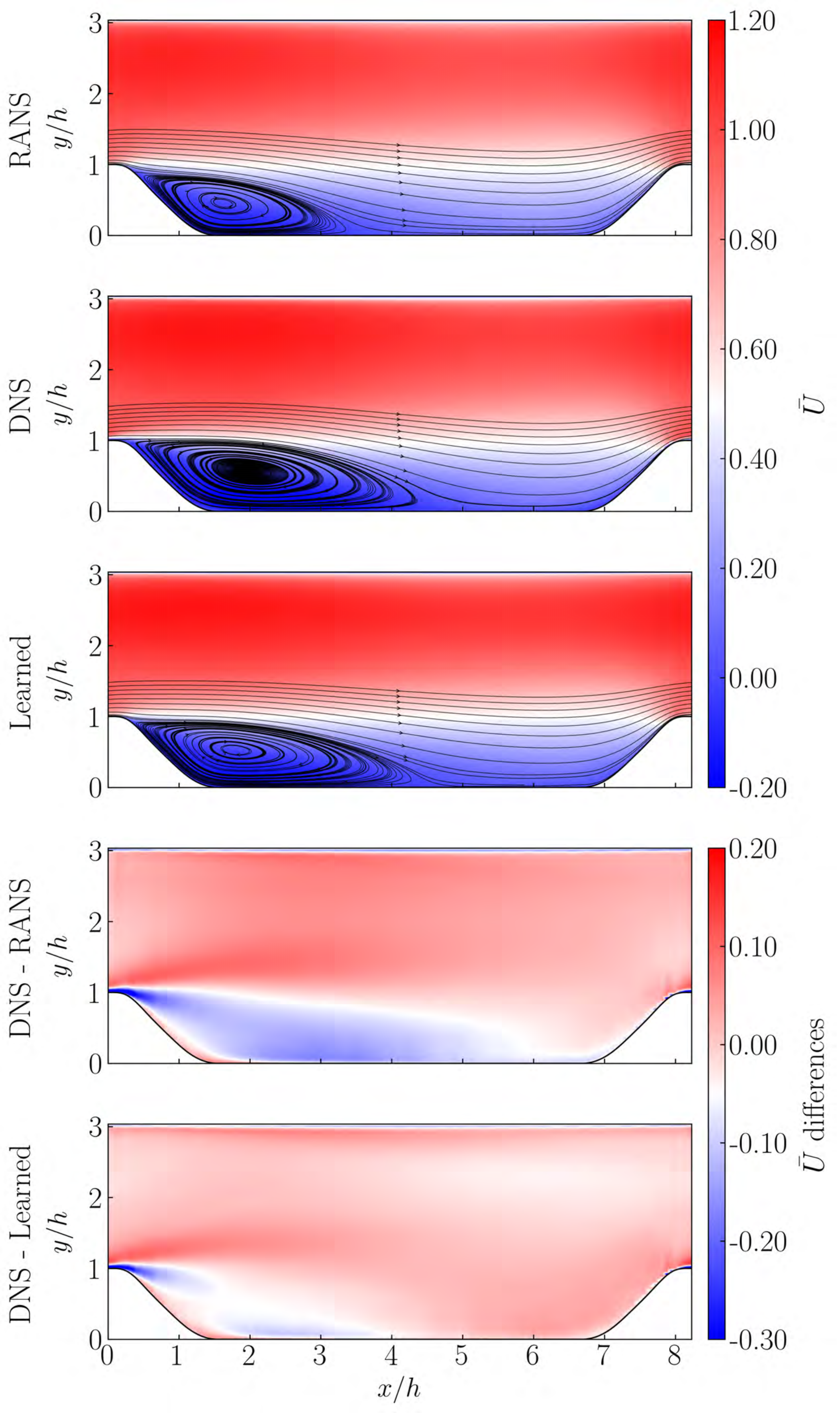}
  }
  \subfigure[]{
  \includegraphics[width=0.45\textwidth]{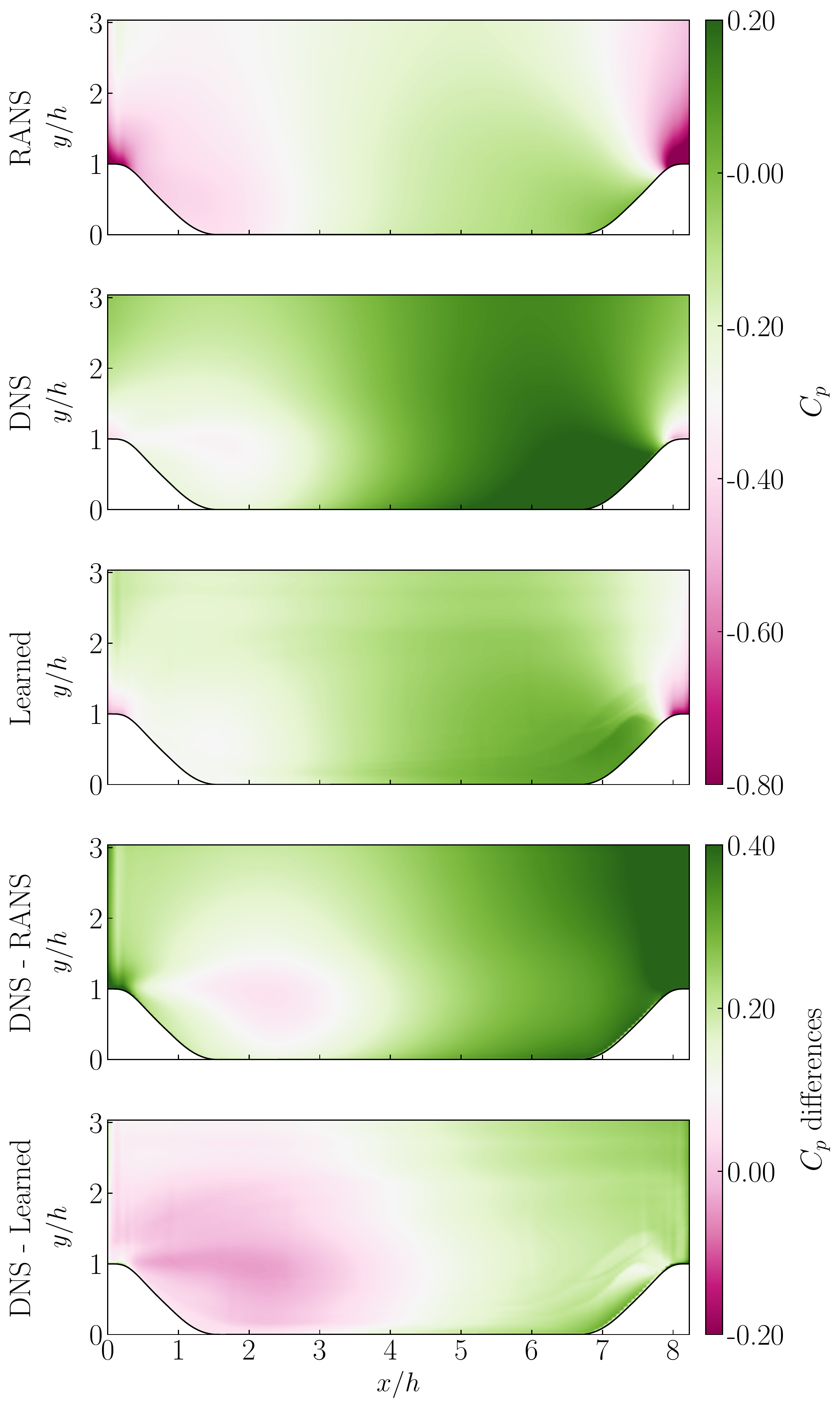}
  } \caption{Comparison of (a) normalized streamwise velocity $\bar{U}$ and (b) pressure coefficient $C_p$ for periodic
  hill training flow ($\alpha=0.8$) using LEVM, DNS, and the learned model, respectively. From top to bottom: baseline
  RANS solutions, time-averaged DNS solutions, the learned model solutions, and the difference between exact data and
  predicted data by LEVM and the learned model, respectively.}
\label{fig:08UP}
\end{figure*}

The improved velocity and pressure fields, compared with the DNS data and the results obtained via standard
$k-\varepsilon$ model, are shown in Fig.~\ref{fig:08UP} (both tow fields are non-dimensional). The streamline resulting
from DNS data in Fig.~\ref{fig:08UP} (a) shows that the baseline RANS model underestimates the size of separation
bubble, revealing that the standard $k-\varepsilon$ model is not capable of precisely capturing the flow separation
feature for this case. The learned model provides more accurate Reynolds stress anisotropy. Therefore, it successfully
enlarges the RANS-predicted separation bubble. An analogous improvement is also reflected by the magnitude of pressure
in Fig.~\ref{fig:08UP} (b). The baseline RANS model over-predicts the pressure magnitude at the upstream and downstream
hillcrest. In addition, it also under-predicts the pressure at the downstream bulk region. By contrast, the result
predicted by the discovered model is more consistent with DNS data. The advantage of the discovered model is more
specifically demonstrated by the velocity and pressure differences. At the upstream hillcrest where flow separation
happens, the relative error of baseline RANS results to DNS results is nearly double larger than the error of learned
flow fields to corresponding DNS flow fields.

\begin{figure*}
  \centering
  \includegraphics[width=0.72\textwidth]{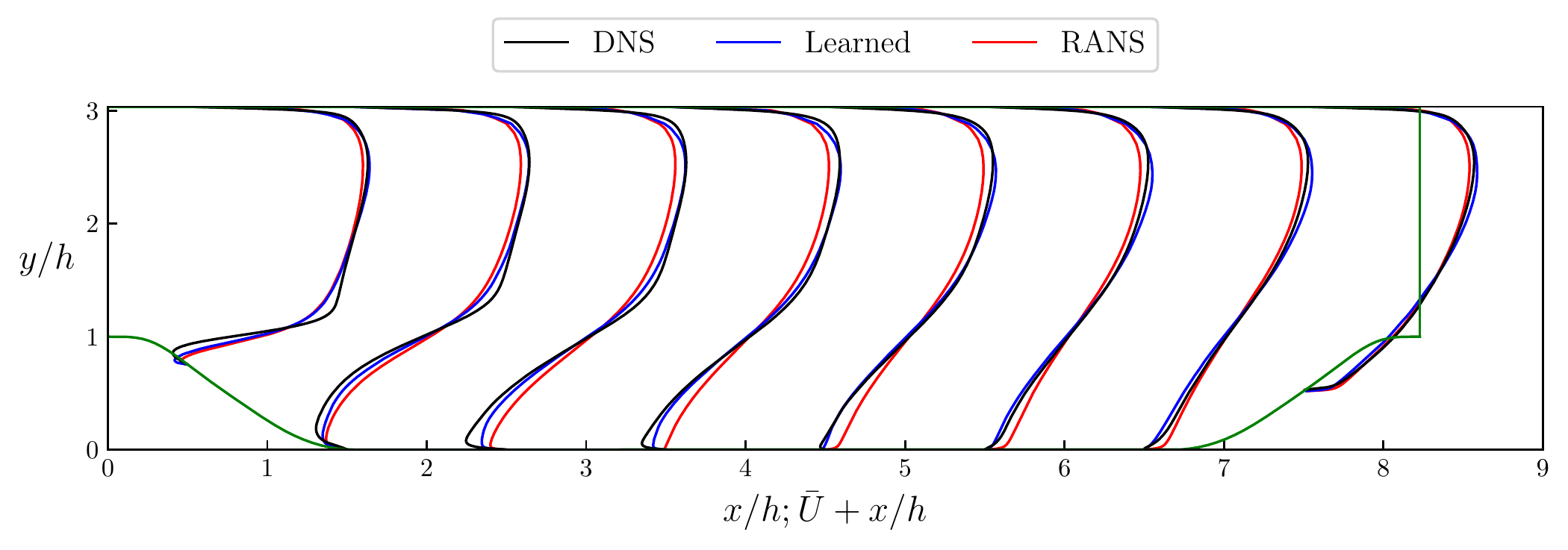}
  \caption{Normalized streamwise velocity profiles from the learned symbolic model for periodic hill training flow
  ($\alpha=0.8$), in comparison with the DNS data as well as the baseline RANS results obtained via $k-\varepsilon$
  model.}
  \label{fig:velProfile}
\end{figure*}

An additional quantitative investigation into the prediction performance of the learned model is described in
Fig.~\ref{fig:velProfile}, in which the predictive velocity profiles are compared against DNS data. The result of the
baseline RANS method shows a large departure from the DNS result. As a comparison, the velocity profiles predicted via
the discovered model demonstrate remarkable improvement over traditional RANS-based results in the whole flow region.

\begin{figure*}
  \centering
  \includegraphics[width=0.98\textwidth]{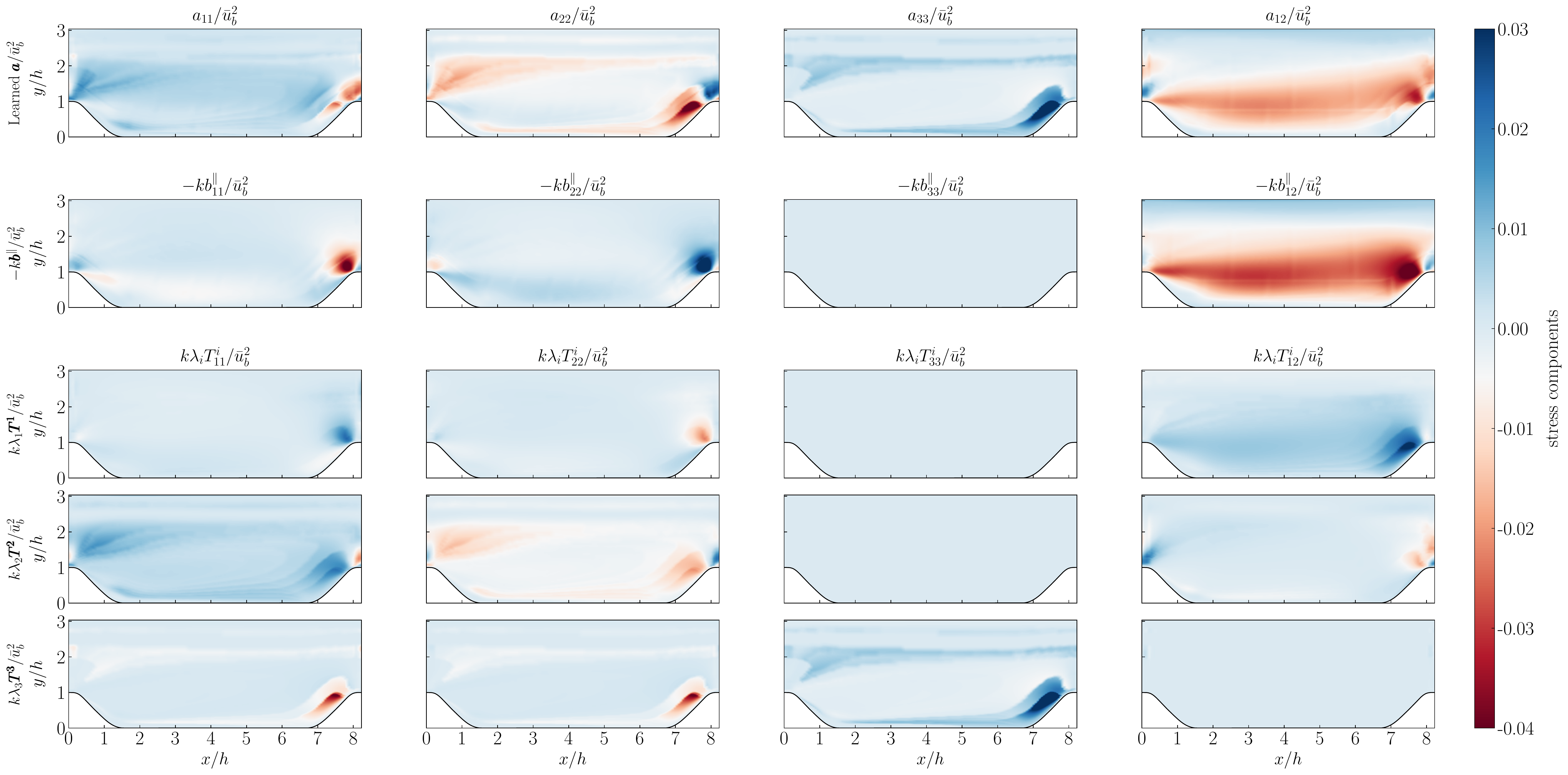}
  \caption{Contribution to each component of the Reynolds stress anisotropy from each tensor basis for the learned
  model as well as the linear part. All results are normalized by the bulk velocity at crest $\bar{u}_b$.}
  \label{fig:08RSCon}
\end{figure*}

By reviewing the RANS governing equation for momentum (Eq.~\ref{eq_mom}), it is straightforward to deduce that the
difficulty inherent in calculating the anisotropic Reynolds stress components $a_{11}$ and/or $a_{12}$ (note that the
flow is two-dimensional) is perhaps the major contributing factor for the incapacity to accurately predict flow
separations with traditional turbulence models. As can be seen in Fig.~\ref{fig:08RS}, while the standard
$k-\varepsilon$ turbulence model generally makes good predictions for shear component $a_{12}$, it cannot capture the
right sign and magnitude of normal component $a_{11}$. The learned model provides relatively better predictions for
these two tensor components. Although the flow past the periodic hill is generally dominated by the shear layer
(especially at the region where flow separation occurs), the deficiency in Reynolds stress anisotropy impairs the
performance of $k-\varepsilon$ turbulence model, rendering the predictions for other quantities of interest, such as
velocity and pressure, inaccurate. To further investigate how the symbolic model contributes to the flow simulation, the
contributions from the tensor basis functions consisting of the learned model $\pmb{b}^{\bot}$ as well as the
iteratively updated linear part $\pmb{b}^{\Vert}$ are presented in Fig.~\ref{fig:08RSCon}. It can be observed that the
linear part $\pmb{b}^{\Vert}$ and the first tensor basis $\pmb{T}^{1}$ dominate the calculation for $a_{12}$, and the
second tensor basis $\pmb{T}^{2}$ is the most important contributor for describing $a_{11}$. Thus the linear part and
the first two bases are the most important ingredients for accurately capturing the flow separation. More precisely, it
is their gradients that determine the calculation of the recirculation region since the momentum equation is solved with
the divergence of Reynolds stress. The third tensor basis $\pmb{T}^{3}$ makes the most important contribution for
modeling $a_{33}$, which enhances the anisotropic property of the predicted stress tensor.

% Fig.~\ref{fig:converge} presents the residual convergence history of the simulation. As previously mentioned, the
% invariants and tensor basis functions fed to the training process are formulated by baseline RANS results. Thus the
% standard $k-\varepsilon$ turbulence model is firstly performed to provide an initial flow field. The discovered symbolic
% model is subsequently applied to the CFD solver and the flow field is iteratively updated until reconvergence. After
% substituting the Reynolds stress, the residual jumps up to a relatively high level and quickly decreases again. The
% convergence is slower than the baseline RANS simulation, but it is still acceptable.

% \begin{figure}
%   \centering
%   \includegraphics[width=0.45\textwidth]{fig/residuals.pdf}
%   \caption{Residual convergence history of the periodic hill training flow ($\alpha=0.8$).}
%   \label{fig:converge}
% \end{figure}

\subsection{Application beyond the training scope}

It has been presented in Section~\ref{sec:08Results} that the model discovered with the present method shows promising
improvements in the predictive accuracy of RANS-based simulations. To evaluate whether the learned model can achieve
similar accuracy improvements for cases outside its training scope, the model in Eqs.~\ref{equ:bbot} and
\ref{equ:coefficients} are applied to three test cases with different geometries, as shown in Fig.~\ref{fig:flowGeo}.
As has been done in previous discussions, the performance of the learned model is assessed by comparing its predictions
against the LEVM and high-fidelity data.

\begin{figure*}
  \centering
  \includegraphics[width=0.98\textwidth]{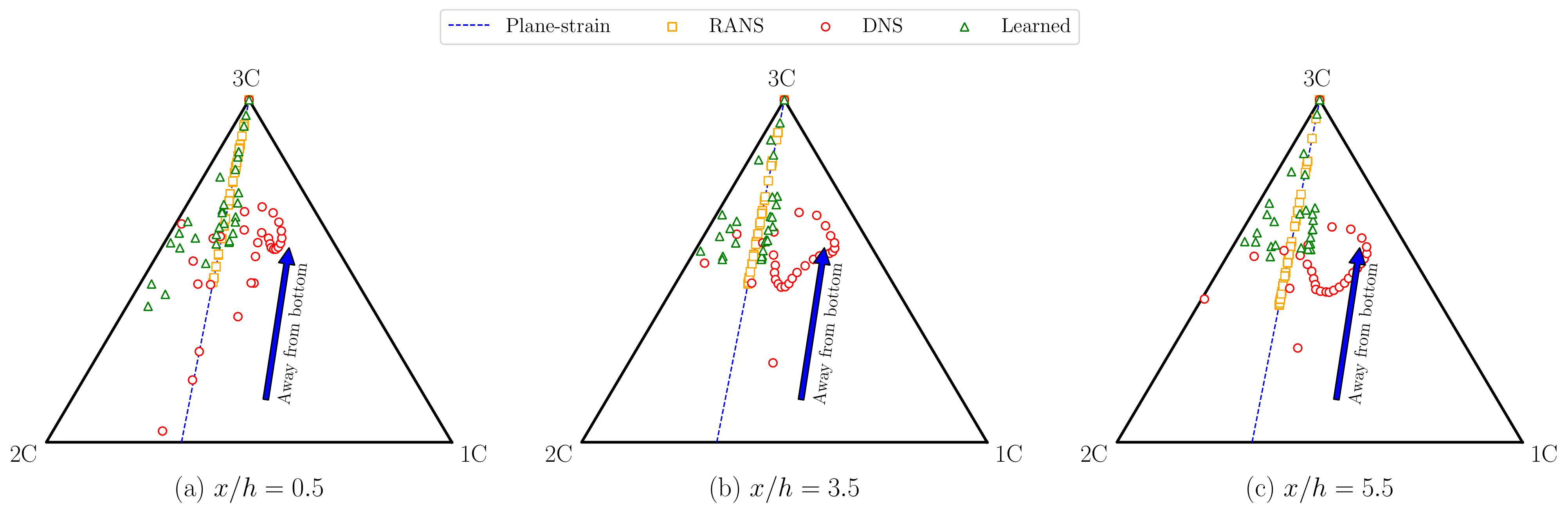}
  \caption{Barycentric map of the predicted Reynolds stress anisotropy for periodic hill test flow ($\alpha=0.5$). The
  learned predictions on three streamwise locations at $x/h=0.5$, $x/h=3.5$, $x/h=5.5$ are compared with the
  corresponding results from high-fidelity DNS simulations and the standard $k-\varepsilon$ RANS turbulence model in
  (a), (b) and (c), respectively.}
  \label{fig:05tri}
\end{figure*}

\begin{figure*}
  \centering
  \includegraphics[width=0.98\textwidth]{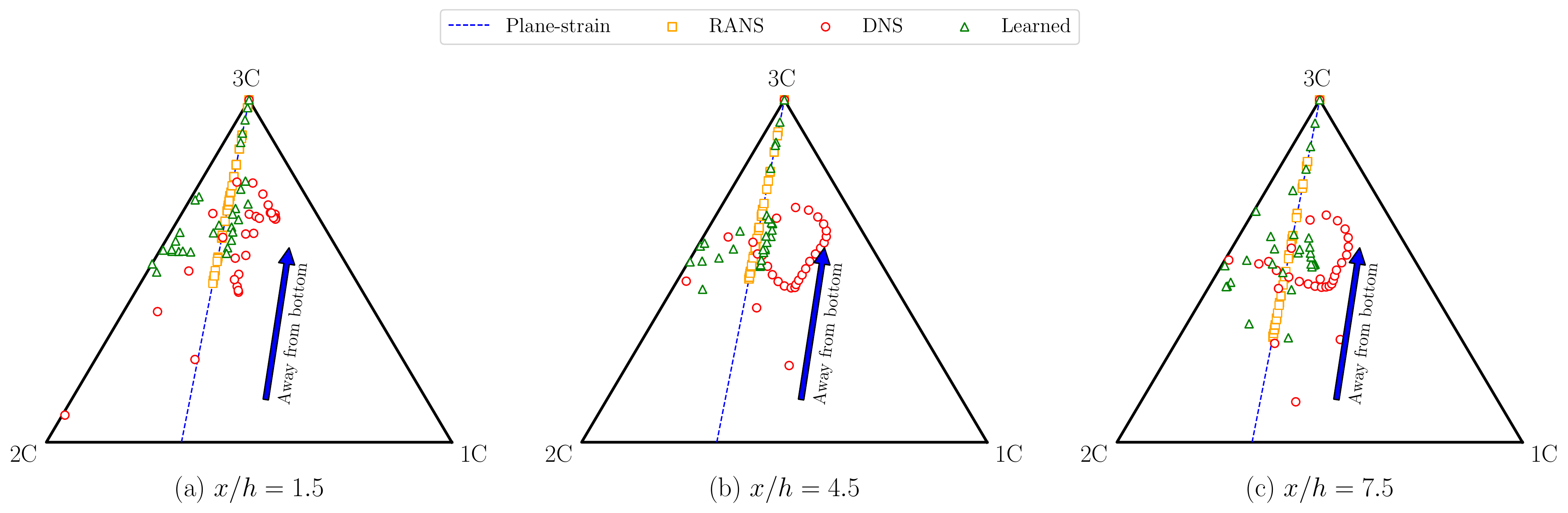}
  \caption{Barycentric map of the predicted Reynolds stress anisotropy for periodic hill test flow ($\alpha=1.0$). The
  learned predictions on three streamwise locations at $x/h=1.5$, $x/h=4.5$, $x/h=7.5$ are compared with the
  corresponding results from high-fidelity DNS simulations and the standard $k-\varepsilon$ RANS turbulence model in
  (a), (b) and (c), respectively.}
  \label{fig:10tri}
\end{figure*}

Although the geometrical configurations of the first two test cases are similar to that of the training case, the flow
behaviors are highly susceptible to the varying hill slope and channel length. Figs.~\ref{fig:05tri} and \ref{fig:10tri}
present barycentric maps of the Reynolds stress anisotropy predicted by the learned model for these two periodic hill
test flows with $\alpha=0.5$ and 1.0, respectively. For each test case, the Reynolds stress anisotropy at three typical
streamwise locations, i.e., the upstream and downstream hill crests as well as the middle flat section, are selected to
demonstrate the prediction performance of the learned model. Similar to the results shown in Fig.~\ref{fig:08tri}, the
realizability condition for anisotropic Reynolds stress tensor is well-maintained by the learned model. In addition, it
clearly shows that the anisotropy predictions at all three cross-sections are markedly improved, albeit not  in complete
accord with the DNS predictions.

\begin{figure*}
  \centering
  \subfigure[]{
  \includegraphics[width=0.72\textwidth]{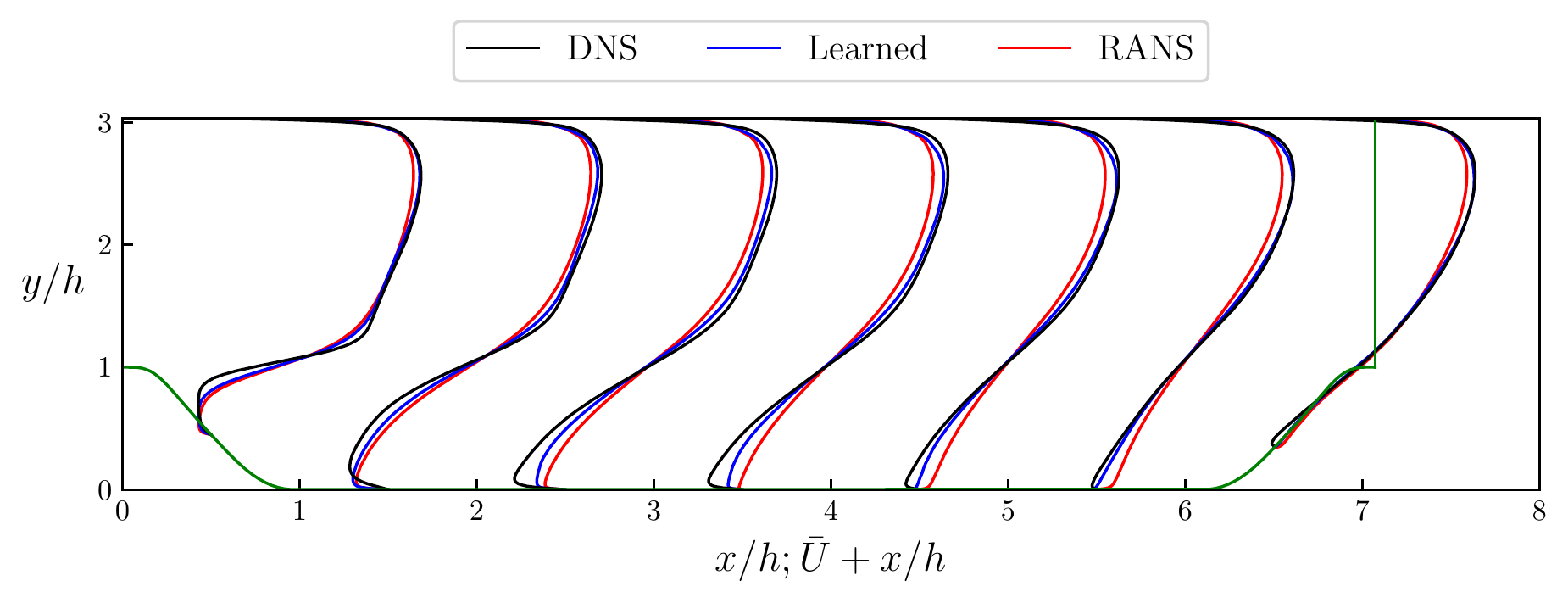}
  }
  \subfigure[]{
  \includegraphics[width=0.72\textwidth]{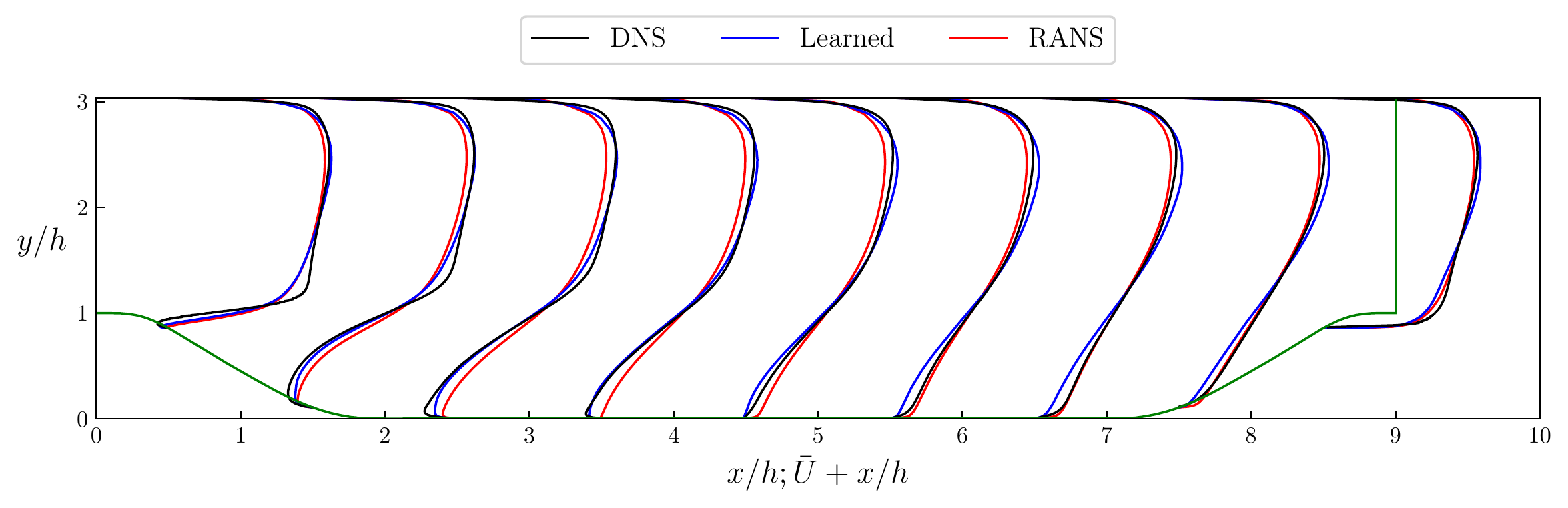}
  } \caption{Normalized streamwise velocity profiles from the learned symbolic model for two periodic hill test cases,
  in comparison with the DNS data as well as the baseline RANS results obtained via the standard $k-\varepsilon$ model.
  (a) $\alpha=0.5$, (b) $\alpha=1.0$}
\label{fig:0510velProfile}
\end{figure*}

\begin{figure*}
  \centering
  \subfigure[]{
  \includegraphics[width=0.45\textwidth]{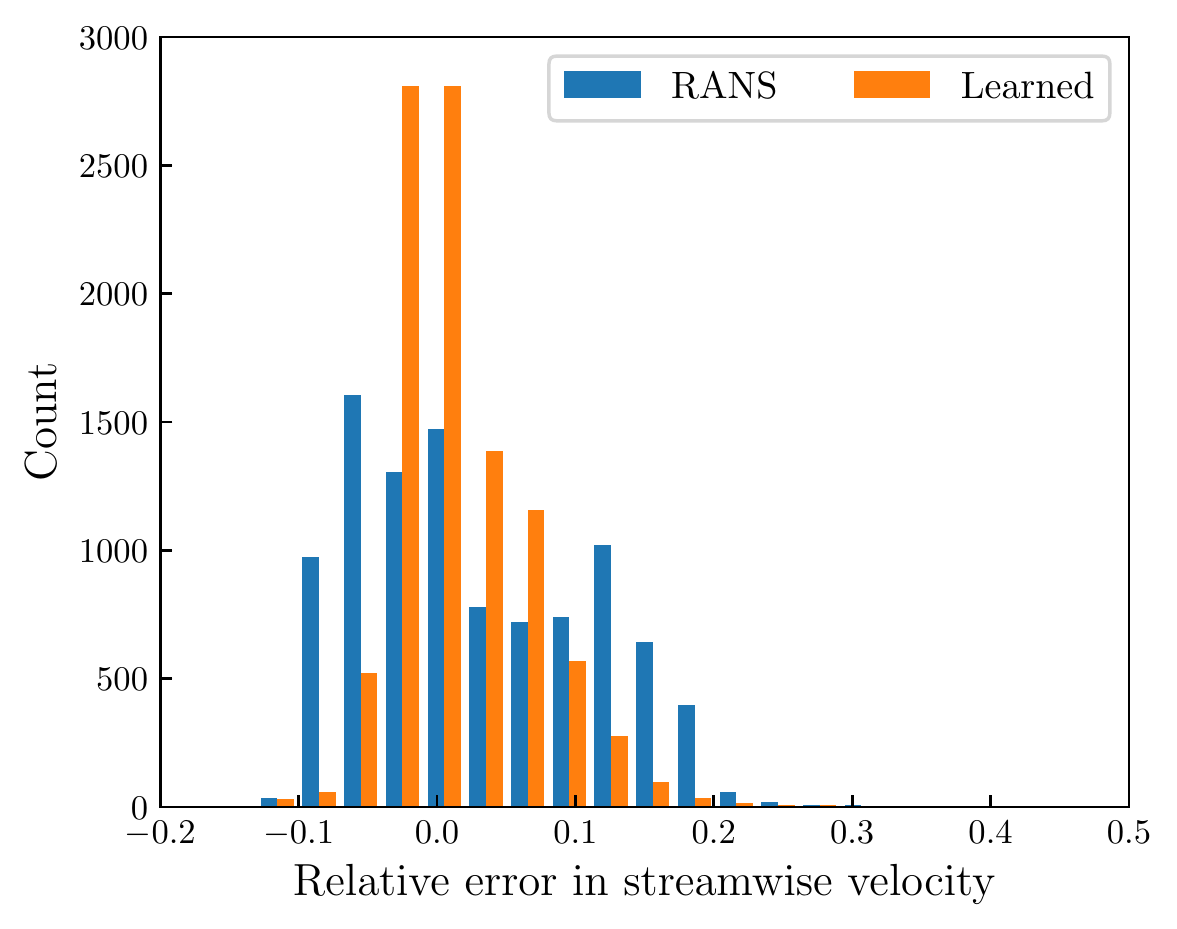}
  }
  \subfigure[]{
  \includegraphics[width=0.45\textwidth]{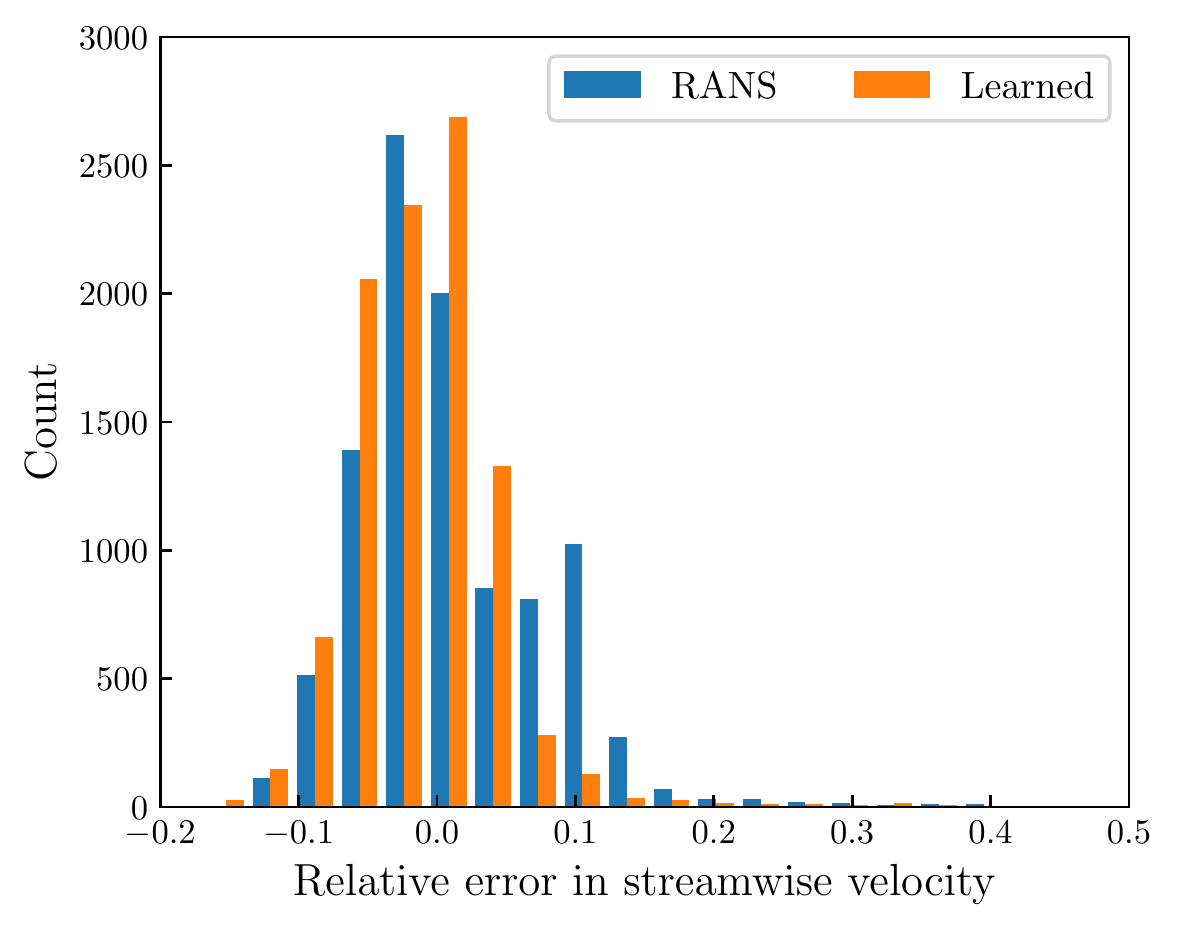}
  } \caption{Error distribution in normalized streamwise velocity for two periodic hill test cases. The relative error
  is calculated by $U_p - U_{DNS}$, where $U_p$ is predicted by the learned model or the standard $k-\varepsilon$ model
  and $U_{DNS}$ indicates the DNS velocity. (a) $\alpha=0.5$, (b) $\alpha=1.0$.}
\label{fig:errorDis}
\end{figure*}

Substituting the corrected Reynolds stress fields into the CFD solver, the velocity improvements over LEVM can be
observed in Fig.~\ref{fig:0510velProfile} for the test cases. While the improvements are not distinguishable for case
$\alpha=0.5$ at the recirculation region, the discovered model indeed makes contributions to improve the accuracy of
flow separation prediction for case $\alpha=1.0$. The observed inaccuracies in case $\alpha=0.5$ can be ascribed to its
larger hill slope, thus the flow separation is more severe than that of the training case. Moreover, the learned model
generally produces better predictions for both test cases at the bulk region. The improvements can also be
quantitatively verified by comparing the overall prediction error in streamwise velocity. Figs.~\ref{fig:errorDis}
presents the distribution of error in streamwise velocity when applying the learned model, compared with the
corresponding results obtained by baseline simulations. For both test cases, it can be observed that the learned model
provides a higher proportion of low relative error than the baseline turbulence model. As the relative error increases,
the number of data points corresponding to the learned model quickly decreases. Despite the learned model could produce
higher errors in part of the computational domain, it generally outperforms the baseline turbulence model.

\begin{figure*}
  \centering
  \includegraphics[width=0.80\textwidth]{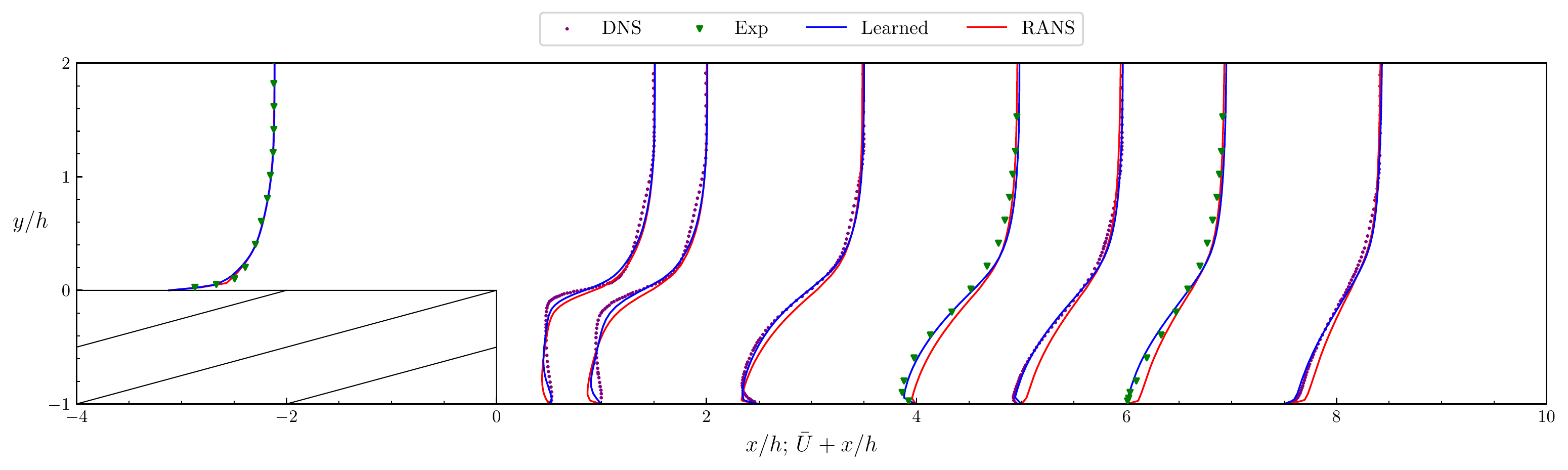}
  \caption{Normalized streamwise velocity profiles (zoomed into the near-step region) from the learned symbolic model
  for backward-facing step test case, in comparison with the reported DNS data as well as the experiment data.}
  \label{fig:bStep5000}
\end{figure*}

The last test case considered here is turbulent flow over a backward-facing step at $Re=5000$. Its geometry is far
removed from the flow scenario under which the symbolic model is discovered. Thus this case would be a more
comprehensive assessment of the generalization performance of the learned model. Since full-field data are not available
for this flow configuration, the performance of discovered model is assessed by comparison with reported DNS data
\citep{jovic1994backward} and experimental data \citep{le_moin_kim_1997}. As expected in Fig.~\ref{fig:bStep5000}, the
velocity profiles predicted by the learned model are much closer to high-fidelity velocity data across the whole domain
(the figure is zoomed into the near-step region due to the limitation of available DNS and experiment data). It should
be emphasized that the symbolic model is only trained on one flow through a periodic hill, but it still achieves
promising improvements for out-of-scope flow configurations.

\subsection{Effect of reward functions}

As introduced in Section~\ref{sec:dso}, the deep symbolic regression method uses a RL algorithm to
train the neural network, thus the choice of the reward function would have a direct effect on the final form of
discovered models. The results discussed above are based on the model using a modified root-mean-square-error function
(Eq.~\ref{equ:reward_rmse}) as its reward function. In this section, another different reward function is utilized
to investigate whether a different symbolic model can be discovered, and more importantly, whether similar accuracy
improvements can be achieved by the new model.

The new reward function is constructed as
\begin{equation}
  -\log(1 + \frac{1}{n}\sum\limits_{i=1}^{n}{\|\widehat{\pmb{b}^{\bot}_i} - \pmb{b}^{\bot}_i\|^2}),\label{equ:reward_log}
\end{equation}
where $n$, $\pmb{b}^{\bot}$ and $\widehat{\pmb{b}^{\bot}}$ keep the same as is in Eq.~\ref{equ:reward_rmse}.
Compared with Eq.~\ref{equ:reward_rmse}, the root-mean-square-error is not normalized in this reward function. With
the use of the logarithmic function, it would be expected that different closure models can be discovered.

Using the same training dataset (i.e., case $\alpha=0.8$ in Fig.~\ref{fig:flowGeo} (a)) and parameters as listed in Tab.~\ref{tab:hyperPara}, the discovered model by taking Eq.~\ref{equ:reward_log} as the reward function is
\begin{widetext}
\begin{equation}
    \begin{aligned}
      \pmb{b}^{\bot} = \frac{\beta}{10}&[(0.8603I_1^2I_2^2 + 0.8603I_1^2I_2 + 0.3014I_1^2 + 0.7206I_1I_2 + 0.5144I_1)\pmb{T}^{1} \\
      &+ (0.5544I_1^2I_2^6 - 0.5544I_1^3I_2^4 - 0.06858I_1^2I_2^4 - 0.5544I_1I_2^2 - I_1I_2 - 0.3865)\pmb{T}^{2} \\
      &+ (-1.328I_1^3I_2^2 + 1.328I_1^2I_2^3 + I_1I_2^2 - 3.289I_1I_2 + 3I_2^2 + 4.624I_2)\pmb{T}^{3}],
    \end{aligned}\label{equ:bbotlog}
\end{equation}
\end{widetext}
where the damping factor $\beta/10$ keeps the same value as in Eq.~\ref{equ:bbot}. For simplifying the discussion
in the following context, the two discovered models in Eqs.~\ref{equ:coefficients} and \ref{equ:bbotlog} are
referred to as model 1 and model 2, respectively.

Comparing the two models discovered by different reward functions, the learned model 2 shows a more complex form, even
though the training of both model 1 and model 2 is carried out using the same dataset and hyper-parameters.
Specifically, for the scalar coefficient corresponding to the same tensor basis, the order of invariants in model 2 is
much higher than that in model 1, which means that there are more multiplying operations when applying model 2. By
contrast, the higher-order terms, such as $0.5544I_1^2I_2^6$, fail to survive in the training evolution of model 1. On
the other hand, it is noted that the symbolic expressions in model 1 contain fewer terms, resulting in a more compact
form of turbulence closure model.

\begin{figure*}
  \centering
  \includegraphics[width=0.98\textwidth]{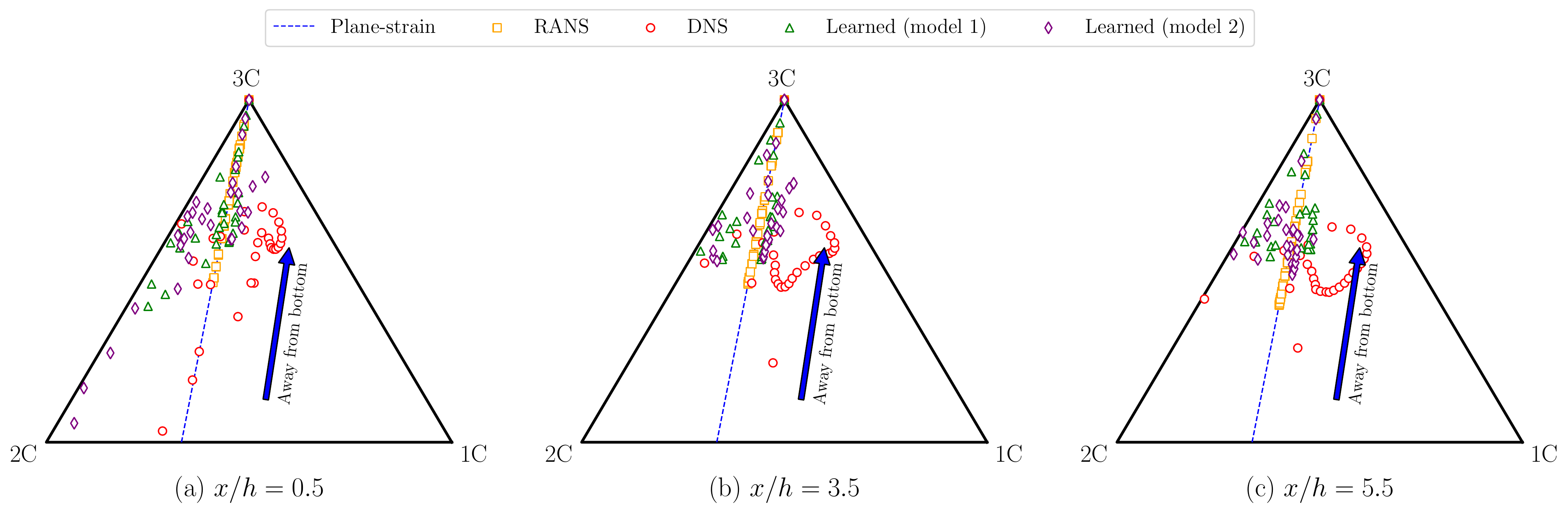}
  \caption{Barycentric map of the Reynolds stress anisotropy predicted by learned model 2
  (Eqs.~\ref{equ:coefficients}) for periodic hill test case $\alpha=0.5$, compared with the corresponding results
  from model 1 (Eqs.~\ref{equ:bbotlog}).}
  \label{fig:05RS12}
\end{figure*}

\begin{figure*}
  \centering
  \includegraphics[width=0.98\textwidth]{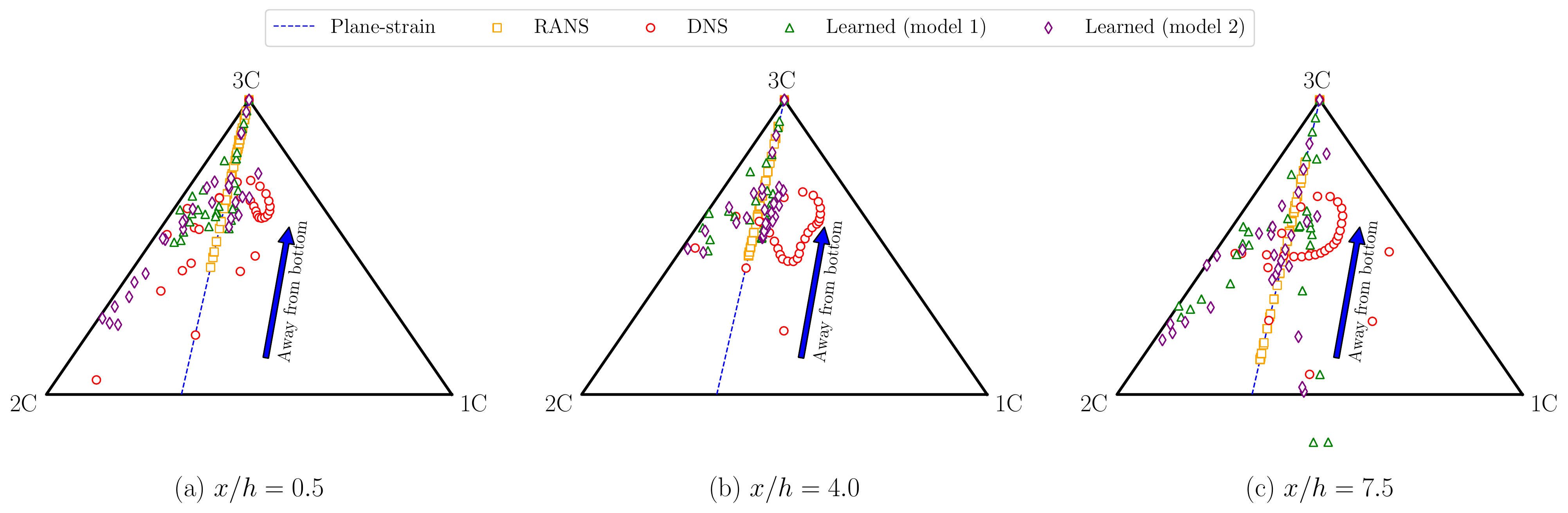}
  \caption{Barycentric map of the Reynolds stress anisotropy predicted by learned model 2
  (Eqs.~\ref{equ:coefficients}) for periodic hill train case $\alpha=0.8$, compared with the corresponding
  results from model 1 (Eqs.~\ref{equ:bbotlog}).}
  \label{fig:08RS12}
\end{figure*}

The next step is to investigate the performance of model 2. Even though model 2 presents a different functional form
from model 1, the use of a physics-based constraint of the tensor basis set could ensure that reasonable accuracy
improvements can be achieved. For brevity, periodic hill cases $\alpha=0.5$ and $\alpha=0.8$ corresponding to the
extrapolated test case and train case are selected to demonstrate the predictive performance of model 2, in comparison
with the results obtained by model 1. As can be seen in Figs.~\ref{fig:05RS12} and \ref{fig:08RS12}, both the two
discovered models can satisfy the realizability requirement. In a sense, model 2 performs even better than model 1 since
there are no outliers in its predictions, as shown in \ref{fig:08RS12} (c). In general, the two learned models
outperform the baseline RANS turbulence model across the whole domain. Although model 2 has a more complex form, it has
no distinct advantages over model 1 in improving the accuracy of Reynolds stress anisotropy.

\begin{figure*}
  \centering
  \subfigure[]{
  \includegraphics[width=0.8\textwidth]{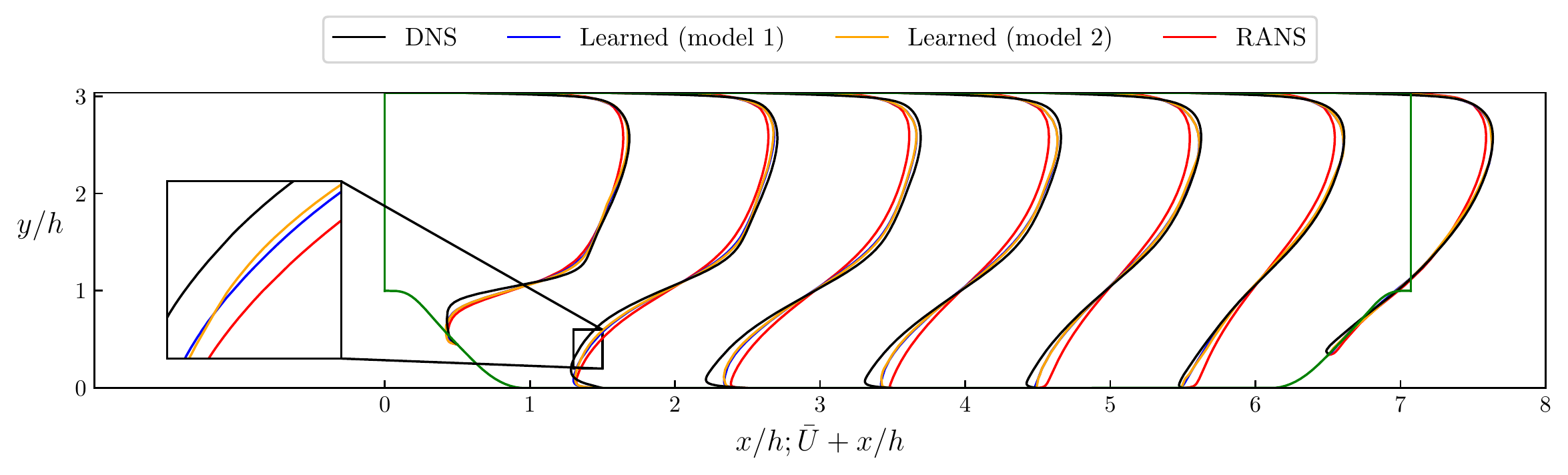}
  }
  \subfigure[]{
  \includegraphics[width=0.8\textwidth]{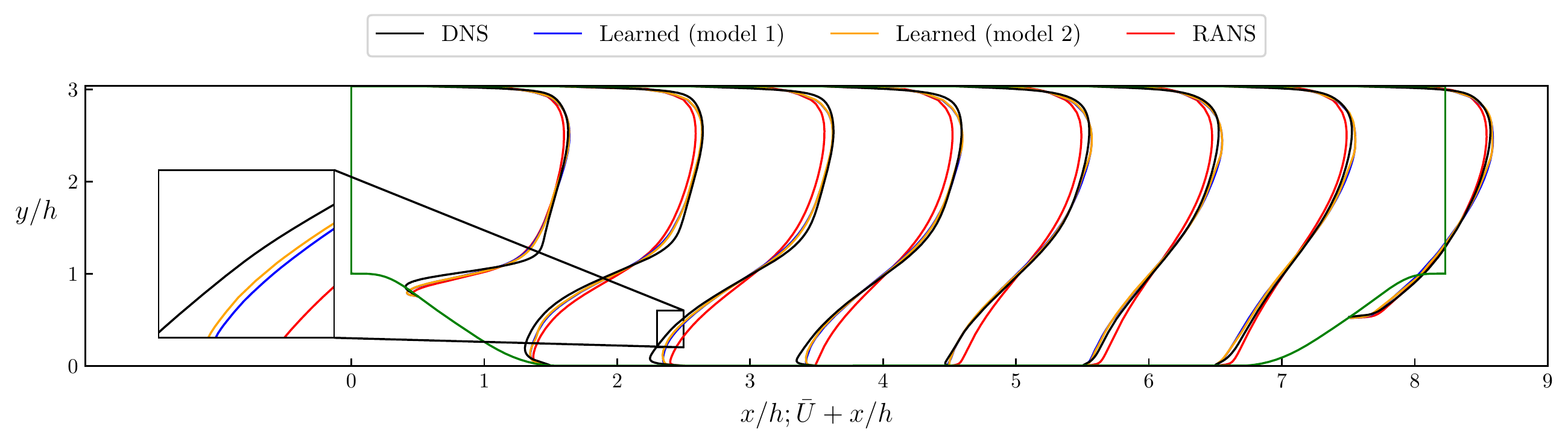}
  } \caption{Normalized streamwise velocity profiles obtained by using the learned model 2, in comparison with the
  corresponding results from model 1. (a) Periodic hill test case $\alpha=0.5$; (b) Periodic hill train case
  $\alpha=0.8$.}
\label{fig:0508velProfile12}
\end{figure*}

Fig.~\ref{fig:0508velProfile12} shows a comparison of streamwise velocity predicted by the two learned models. It is
noted that the two discovered models result in strikingly close velocity fields, even though model 2 bears little
resemblance to model 1. This scenario can be ascribed to the statistical error associated with velocity and Reynolds
stress. Specifically, the mean velocity is a first-order statistic, whereas the Reynolds stress is a second-order
statistic that is less converged, as has been reported by \citet{THOMPSON20161}. Compared with baseline RANS
predictions, both two discovered models can achieve promising accuracy improvements for all flow cases, even if the
corresponding configuration is outside the training scope.

\section{Conclusions}\label{sec:conclusions}

In this work, a data-driven RANS turbulence modeling approach based on the combination of deep learning and symbolic
regression techniques is proposed. This approach leverages the representational capability of deep learning to search
the symbolic space for generating interpretable expressions. A risk-seeking RL algorithm is leveraged to train the LSTM
neural network so as to maximize the best sampled expressions. The resultant turbulence closures are explicitly given in
the form of algebraic polynomials via the embedded invariants and tensor basis functions, thus allowing direct
functional inference and achieving promising Galilean invariance properties. In addition, with the use of present
method, it is straightforward to implement the discovered model equations into existing RANS solvers, without the need
to deploy a deep learning environment for new data-driven simulations.

The performance of the proposed approach is validated by three canonical flows that differ in geometrical
configurations. Although the training dataset consists of only one flow through a periodically constricted channel, the
learned turbulence model, on the whole, demonstrates promising accuracy improvements compared with LEVM for all test
flows, even for extrapolated cases that could hold flow features not seen during the training process. The algebraic
form renders the discovered model more realizable in practical RANS simulations. Additionally, the reward function plays
an important role in discovering the symbolic turbulence model since the neural network is trained by RL algorithms. In
the context of this study, two reward functions are employed. While the two discovered turbulence models differ greatly
in function forms, their predictions show a certain degree of similarities, especially for the low-order statistics.

Despite the relative simplicity of the selected RANS simulations, the experience and insights gained from this work shed
a light on the future development of interpretable data-driven turbulence models. One of the most challenging issues is
developing a reliable and interpretable model that can present universally good performance for complex industrial flow
simulations. The model performance could be enslaved by the policy used in RL training \citep{novati2021automating}.
Another issue is about the relatively high computation cost regarding the constant optimization in DSR method, or more
generally, the training. Fuelled by the advances of symbolic regression methods and RL algorithms in ML community, it is
anticipated that more comprehensive frameworks can be constructed for turbulence modeling research. By integrating the
RANS simulations with RL-based training process, the reward can be defined by other quantities of interest, so that the
discovered model would be more consistent with physical observations \citep{novati2021automating}. In addition, by
extending RL with special types of neural networks, a non-local constitutive models could be developed for improving the
performance of data-driven turbulence models \citep{ZHOU2021113927, KURZ2023109094}.

\section{Data availability}

The relevant code and data used in this project are publicly available at \url{https://github.com/thw1021/DSRRANS}. The
DSR method is implemented by \citet{petersen2020deep} using deep learning library TensorFlow
\citep{Abadi_TensorFlow_Large-scale_machine_2015}, and the CFD simulations are carried out by using OpenFOAM
\citep{doi:10.1063/1.168744}.

\section{Acknowledgements}

This work is supported by the National Key Research and Development Program (Grant Nos. 2019YFE0192600 and
2019YFB1503700), Natural Science Foundation of China (Grant No. 52006098), and Priority Academic Program Development of
Jiangsu Higher Education Institutions. Y. Wang acknowledges the support of the Natural Science Foundation of China
(Grant Nos. 11902153 and 12272178), the Research Fund of State Key Laboratory of Mechanics and Control of Mechanical
Structures (Grant No. MCMS-I-0122G01) and Key Laboratory of Computational Aerodynamics, AVIC Aerodynamics Research
Institute.

% \begin{appendices}
\appendix

\section{Brief overview of LSTM neural network}\label{appendix:lstm}

\begin{figure*}
  \centering
  \includegraphics[width=0.96\textwidth]{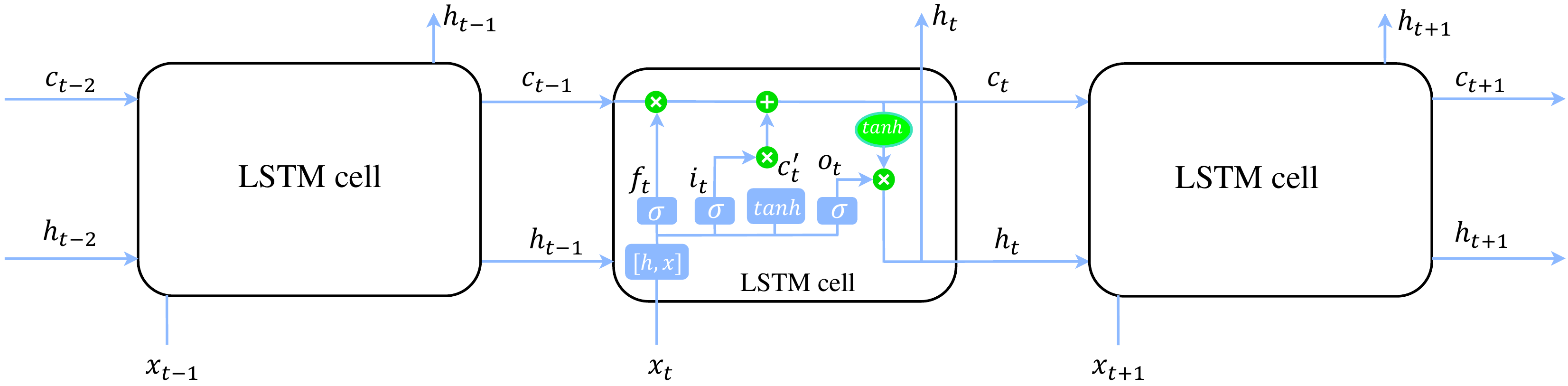}
  \caption{Architecture of the LSTM network.}
  \label{fig:lstm}
\end{figure*}

The LSTM network is a special variant of RNN. It is designed to process data sequences and utilizes its internal memory
to learn and harness information relative to what it has seen so far. As a consequence, the network predictions are not
only determined by the current input but also conditionally dependent on the recent input sequence. A schematic of the
LSTM network is shown in Fig.~\ref{fig:lstm}. A typical LSTM cell contains three gates: the input gate $i$, output
gate $o$ and forget gate $f$. The cell input and output are given by $x$ and $h$, respectively.

Compared to the standard RNN, the LSTM network introduces an additional mechanism to carry information, i.e., the cell
state $c$, across many timesteps. The basic equations involved in the forward computation of LSTM networks are given as follows
\begin{equation}
  \begin{aligned}
    f_t &=\sigma\left(W_f \cdot\left[h_{t-1}, x_t\right]+b_f\right), \\
    i_t &=\sigma\left(W_i \cdot\left[h_{t-1}, x_t\right]+b_i\right), \\
    c'_t &=\tanh \left(W_c \cdot\left[h_{t-1}, x_t\right]+b_c\right), \\
    c_t &=f_t * c_{t-1}+i_t * c'_t, \\
    o_t &=\sigma\left(W_o \cdot\left[h_{t-1}, x_t\right]+b_o\right), \\
    h_t &=o_t * \tanh \left(c_t\right),
  \end{aligned}
\end{equation}
where $\sigma$ is the sigmoid activation function, $W$ and $b$ are the weights and biases, respectively. As indicated in
Fig~\ref{fig:lstm}, the cell state $c$ is combined with the input connection and the recurrent connection, and it will
work together with the three gates to affect the next cell state by adding or removing information. Conceptually, the
carry dataflow keeps reading and writing from memory, thus allowing past information to be injected at a later time.
This mechanism is the key ingredient that ensures the LSTM network does not suffer from vanishing or exploding
gradients.

\section{RL with the risk-seeking policy gradient algorithm}\label{appendix:rspg}

\begin{figure}
  \centering
  \includegraphics[width=0.45\textwidth]{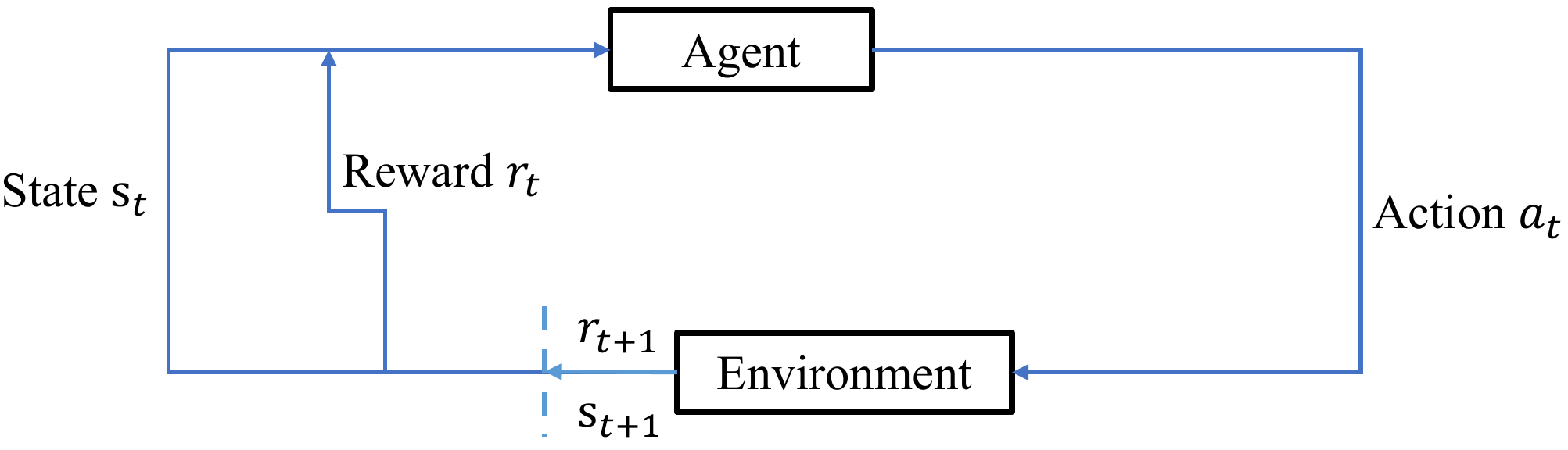}
  \caption{A schematic of the RL framework.}
  \label{fig:rl}
\end{figure}

RL is one of the main branches of machine learning. It is concerned with how to learn through trial and error from
environmental feedback so as to maximize a numerical reward signal. A simplified overview of the RL framework is
presented in Fig.~\ref{fig:rl}. As illustrated, there are two core components, the agent and environment, in RL. The
environment represents the problem and the agent attempts to find the solution to the problem. In general, the learning
agent interacts with the environment through three channels: the state, action and reward. The agent perceives the state
of its environment and chooses actions to affect the environment. The environment then reacts to the actions by
producing a reward signal. After finishing training, the agent is expected to perfectly map situations to actions so
that the reward signal can be maximized.

In the present work, the risk-seeking policy gradient algorithm is employed to train the LSTM neural networks to produce
better-fitting symbolic expressions. The core idea of the risk-seeking policy gradient can be expressed as
\begin{equation}
  J(\theta;\epsilon)=\mathbb{E}_{\tau\sim p(\tau\mid\theta)}\left[R(\tau)\geq R_{\epsilon}(\theta)\right],
\end{equation}
where $R_{\epsilon}(\theta)$ denotes the $(1-\epsilon)$-quantile of the reward distribution produced by the current
policy. Therefore the learning objective $J(\theta;\epsilon)$ is to maximize the reward of the top $\epsilon$ fraction
of samples. The samples below the threshold will not involve in the training process. This strategy has been proven to
be instrumental in improving the performance of sampled symbolic expressions. In the context of this work, the
distribution over mathematical expressions $p(\tau\mid\theta)$ is like a policy. The algorithm agent samples new tokens
when the parent and sibling inputs are observed. During every episode, the agent generates a sequence of expressions,
and the reward value is then calculated by the reward function.

% \end{appendices}
% \bibliographystyle{unsrt}
\bibliography{aipsamp}% Produces the bibliography via BibTeX.

\end{document}